\documentclass[12pt]{article}
\usepackage{ifpdf}
\usepackage{graphicx,amssymb,amsmath,lineno}
\usepackage{enumitem}
\usepackage{bm}
\usepackage[percent]{overpic}
\usepackage{setspace}
\usepackage{scalerel}
\usepackage{tcolorbox}

\usepackage{fancyvrb}
\definecolor{commentcolor}{rgb}{0.5,0.5,0.5}
\definecolor{keywordcolor}{rgb}{0.0,0.0,1.0}
\definecolor{stringcolor}{rgb}{0.0,0.5,0.0}

\usepackage{newfloat}
\usepackage{caption}
\DeclareFloatingEnvironment[fileext=lol,listname={List of Code Listings},name=Listing]{codelisting}

\usepackage[
fencedCode,                 % recognise ``` fences (default true)
tightLists = false,         % keep enumitem compatible
cacheDir   = .mdcache       % where auxiliary files go
]{markdown}

\markdownSetup{
	rendererPrototypes = {
		inputVerbatim   = {\lstinputlisting[basicstyle=\ttfamily\small,frame=none,xleftmargin=0pt]{#1}},
		inputFencedCode = {\lstinputlisting[basicstyle=\ttfamily\small,frame=none,breaklines=true,xleftmargin=0pt]{#1}},
	},
}

\usepackage[T1]{fontenc}

\usepackage{fancyhdr}
\usepackage{listings}
\usepackage{xcolor} % For customizing colors
\usepackage{csquotes}

\MakeOuterQuote{"}

\usepackage[margin=1in, paperwidth=8.5in, paperheight=11in]{geometry}

\usepackage{todonotes}

\ifpdf
\usepackage[%
  pdftitle={Instructions for use of the document class
    elsart},%
    pdfstartview=FitH,%
  bookmarks=true,%
  bookmarksopen=true,%
  breaklinks=true,%
  colorlinks=true,%
  linkcolor=blue,anchorcolor=blue,%
  citecolor=blue,filecolor=blue,%
  menucolor=blue,%
  urlcolor=blue]{hyperref}
\else
\usepackage[%
  breaklinks=true,%
  colorlinks=true,%
  linkcolor=blue,anchorcolor=blue,%
  citecolor=blue,filecolor=blue,%
  menucolor=blue,%
  urlcolor=blue]{hyperref}
\fi
\usepackage{tabularx}
\usepackage{natbib}

\lstdefinestyle{cppstyle}{
	language=C++,
	basicstyle=\ttfamily\footnotesize,
	keywordstyle=\color{blue!70!black},
	commentstyle=\color{gray}\itshape,
	stringstyle=\color{green!50!black},
	identifierstyle=\color{black},
	numbers=none,
	backgroundcolor=\color{white},
	tabsize=4,
	showspaces=false,
	showstringspaces=false,
	frame=single,
	framesep=6pt,
	breaklines=true,
	breakatwhitespace=true,
	captionpos=b,
	morekeywords={uint,vec3,Context,PlantArchitecture,ShootParameters},
}

\lstdefinestyle{htmlstyle}{
	language=XML,
	basicstyle=\ttfamily\footnotesize,
	sensitive=true,
	morecomment=[s]{<!--}{-->},
	morestring=[b]",
	morestring=[b]',
	tagstyle=\color{blue!70!black},
	keywordstyle=\color{blue!70!black},
	commentstyle=\color{gray}\itshape,
	stringstyle=\color{green!50!black},
	numbers=none,
	tabsize=2,
	showspaces=false,
	showstringspaces=false,
	frame=single,
	framesep=2pt,
	breaklines=true,
	backgroundcolor=\color{white},
	captionpos=b,
}

\lstdefinestyle{pythonstyle}{
	language=Python,
	basicstyle=\ttfamily\footnotesize,
	keywordstyle=\color{blue!70!black},
	commentstyle=\color{gray}\itshape,
	stringstyle=\color{green!50!black},
	identifierstyle=\color{black},
	numbers=none,
	backgroundcolor=\color{white},
	tabsize=4,
	showspaces=false,
	showstringspaces=false,
	frame=single,
	framesep=6pt,
	breaklines=true,
	breakatwhitespace=true,
	captionpos=b,
	morekeywords={Context,PlantArchitecture,vec3},
}

\title{
	\vspace{3cm} % Adjust vertical space
	\textbf{\Huge A generalized framework for procedural generation of three-dimensional static and dynamic plant model geometries} \\
	\vspace{3cm}
	{\huge Technical Report}\\
	{\Large December, 2025}\\
	\vspace{2cm}
	{\LARGE Brian N. Bailey}\\
	{\Large Department of Plant Sciences\\
		University of California, Davis}\\
	{\Large bnbailey@ucdavis.edu}\\
	\vspace{4cm}
}
\date{}

\begin{document}
	
	% Title Page
	\maketitle
	\thispagestyle{empty}
	\newpage

%\pagestyle{plain}

%\author{Brian N. Bailey \corref{cor}\fnref{ucdps}}\ead{bnbailey@ucdavis.edu}

%\cortext[cor]{Corresponding author}
%\address[ucdps]{Department of Plant Sciences, University of California, Davis, Davis, CA USA}

%\linenumbers
%\doublespacing

\hrule
\hrule

\section*{Abstract}

This work presents a new framework for procedural generation of dynamic 3D plant model geometries, which has been implemented in the Helios modeling system. Key goals of this work were to develop a model that 1) has a generalized set of parameters that are conserved across species, which are botanically-consistent and readily measurable; 2) significantly reduces the time and effort needed to create photorealistic, dynamically evolving plant models; 3) allows for encoding of the entire plant structure into a character-based representation that can integrated with machine learning models, and 4) includes realistic and computationally efficient collision physics. A model framework that satisfies these specifications is presented in this report. The model was implemented in the Helios C++ and PyHelios Python frameworks, which are open-source libraries that can be used to generate 3D plant geometries based on this model.

\vspace{1cm}
\hrule
\hrule

\vspace{8cm}
\noindent Revision History:

\noindent Version 1: December 2025

\vspace{3cm}
\noindent\copyright{} 2025, the author.

\newpage

\tableofcontents

\listoftables

\newpage

\section{Introduction}\label{intro}

% Introduction of varous types of 3D plant models (computer graphics, FSPM, synthetic data), and motivate that they need resolved 3D architectural inputs
Since the advent of 3D computer graphics, there has been a need for detailed 3D models of plants in order to render landscape scenes. These models consist of a mesh of many small planar elements such as triangles that together form the resolved geometry of the plant. Plants are difficult to represent computationally, and thus also to render, because they usually consist of thousands of small discrete leaves and branches, which are connected through a complex and stochastic growth topology.

% Review prior approaches for generating 3D plant models (reconstructions from measurement; static vs. dynamic; Xfrog vs. L-systems; Weber-Penn )
Many methodologies have been proposed for representing plant topology for generation of 3D models (primarily trees), with the 1980s being among the most active periods for new model development. Early modelers considered plants as fractals and borrowed concepts from L-systems \citep{Lindenmayer71} to recursively generate plant topology \citep[e.g.,][]{Smith84,Bloomenthal85}. These approaches used "grammars", or symbolic representations of plant structure that are iteratively re-written based on rules that govern how a base structure evolves to form the plant topology. This general rule-based approach that evolved from the original ideas of L-systems may be considered "bottom-up" methods in that they start from an elemental base unit and a set of rules, which reproduces to generate an emergent large-scale structure that was unspecified initially. In contrast, "top-down" methods have also been proposed that are primarily based on large-scale parameters (e.g., plant height, branch length, number of child branches per parent branch), then the overall topology is generated to satisfy the specified parameters \citep{Oppenheimer86,Weber95,Lintermann98}. The bottom-up approach is typically more common in applications where a dynamically evolving (i.e., growing) plant is desired, whereas a top-down approach may be more likely to be used where static models at a given instant in time are sufficient.

Further evolution in plant model algorithms brought a desire for models that are consistent with the actual biology of growing plants, which later led to the use of 3D plant models in botanical research. The models of \citet{deReffye88} (and to a more limited extent \citet{Aono84} prior) focused on incorporating botanical knowledge into their models to improve visual realism. While these models sought to incorporate more biological realism, the end goal was still generally to render more realistic plants and not necessarily to use them as a tool for biological research. 

The work of \citet{Prusinkiewicz90}, which built on prior work adapting the original ideas of L-systems to 3D plant models \citep[e.g.,][]{Smith84,Oppenheimer86}, enabled application of L-systems-based models in biological research examining dynamic plant growth and development. The approach of \citet{Prusinkiewicz90} made chronology explicit by linking the rule-based iteration with a developmental time step, which made it a natural tool for modeling plant development. This eventually led to a new field of botanical research based on application of functional-structural plant models (FSPMs) to quantitatively study plant development and environmental interactions using 3D modeling tools \citep{Sievanen00,Godin05}. L-systems-based approaches are by far the most widely used models for generating plant geometry in modern FSPMs, and form the basis of modeling frameworks such as GroIMP \citep{Hemmerling08} and OpenAlea \citep{Pradal08}, among others. Inclusion of L-systems models within an FSPM allows biophysical models to influence developmental behavior within an architectural model, thus constraining them based on available light and carbon \citep{Allen05,Zhu18}.

An additional class of models was proposed by \citet{Runions07} and \citet{Palublicki09} based on "space colonization", whereby branch trajectories and bud fates are chosen in order to fill empty space. This approach breaks symmetry inherent in L-systems, and generally leads to more realistic looking plants. This approach has most commonly been used for static geometry generation rather than being incorporated into dynamic FSPMs, as it does not incorporate chronology needed to model growth. It is also computationally costly, as it must perform many expensive operations during plant generation.

% Limitations of L-Systems and current gaps
% Key limitations:
% 1. Difficult to calibrate from data
% 2. Lack some flexibility 
% 3. Can lack realism due to the high symmetry imposed by growth rules
Current 3D plant architectural models present some limitations in their applications toward the next generation of questions sought to be answered by researchers using FSPMs. One issue is that they are traditionally difficult to calibrate from field data, which is especially true for bottom-up models like L-systems. Their bottom-up nature complicates manual parameter measurement or inferring them indirectly from sensing data. Parameters are often tuned such that they "look about right" in comparison with photographs. Recent work has attempted to automatically calibrate L-systems models from sensing data, but they generally fall short of full calibration and instead assume growth rules and prune or alter the plant to match a measured crown envelope \citep[e.g.,][]{Stava14,Guo18,Liu21}. Bottom-up models like L-systems can also be limiting in terms of their flexibility and level of realism due to the inherent symmetry that such a model exhibits. Although developmental behavior can be randomized, the recursive application of growth rules tends to produce plants with high degrees of hierarchical symmetry, which may reduce visual realism. The string-based encoding used by L-systems models also generally lacks the full suite of information needed to uniquely encode a particular plant geometry, limiting its utility to some degree in automated parameter calibration algorithms. 

% Goals of the present model framework
This work presents a new 3D plant architectural modeling framework with the following novel features:

\begin{enumerate}
	\item The model evolves dynamically in time "bottom-up", but users specify a set of parameters rather than rules that determine emergent plant form.
	\item The model utilizes a single parameter set that can be varied to produce a very wide range of angiosperm plant species.
	\item The model has high flexibility to allow users to specify custom organ models, write custom functions to override behavior, and manually manipulate plant structure.
	\item The model optionally utilizes an efficient collision detection methodology to include self-avoidance during growth, solid obstacle avoidance, and constrained growth (e.g., from a trellis).
	\item The structure of a given plant can be encoded into an XML file, which can be used to interface with external programs through file I/O.
\end{enumerate}

The model code is implemented within the open-source Helios C++ modeling framework (\url{https://www.github.com/PlantSimulationLab/Helios}), and can produce highly realistic 3D plant models that are fully integrated with the suite of other biophysical models within Helios. It has also been incorporated within the PyHelios python programming interface for Helios (\url{https://www.github.com/PlantSimulationLab/PyHelios}). 

\section{Methodology}

\subsection{Overall model structure}

\begin{figure}
	\centering
	\includegraphics[width=0.5\textwidth]{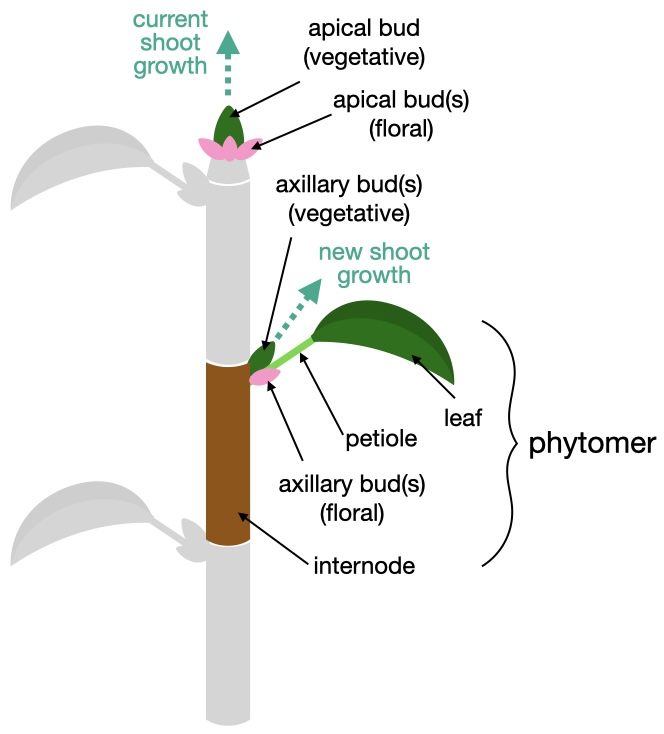}
	\caption{Schematic illustration of the phytomer, the fundamental unit of the plant structure, as a sub-unit of a shoot. The phytomer consists of an internode, one or more petioles connected at the distal end of the internode, and one or more leaves/leaflets connected at the distal end of the petiole. The base of the petiole may have one or more axillary vegetative buds that can produce an inflorescence, which consists of a peduncle, and one or more flowers that can potentially develop into fruit. The shoot consists of one or more phytomers connected end-to-end in series. An apical vegetative bud at the tip of the shoot potentially spawns new phytomers, thus increasing the length of the shoot. Axillary vegetative buds located at the base of the petiole can also spawn new lateral shoots.}
	\label{fig:phytomerschematic}
\end{figure}

Like many other 3D procedural plant growth models, the fundamental geometric unit of the plant is the "phytomer", which consists of an internode with one or more petioles attached at the distal end, where each petiole has one or more leaves/leaflets (Fig. \ref{fig:phytomerschematic}). At the base of the petiole, there may be one or more axillary floral buds that can turn into an inflorescence structure consisting of a peduncle from which there may be one or more flowers that can develop into fruit. At the base of the petiole, there may also be one or more axillary vegetative buds that can turn into a child shoot.

A shoot consists of one or more phytomers with internodes place end-to-end. At the tip of the shoot (tip of most distal phytomer), there may be an apical vegetative bud that can spawn new phytomers. In the case of perennial plants, there may also be one or more apical floral buds located at the shoot tip that can turn into flowers that may develop into fruit.

\begin{figure}
	\centering
	\includegraphics[width=\linewidth]{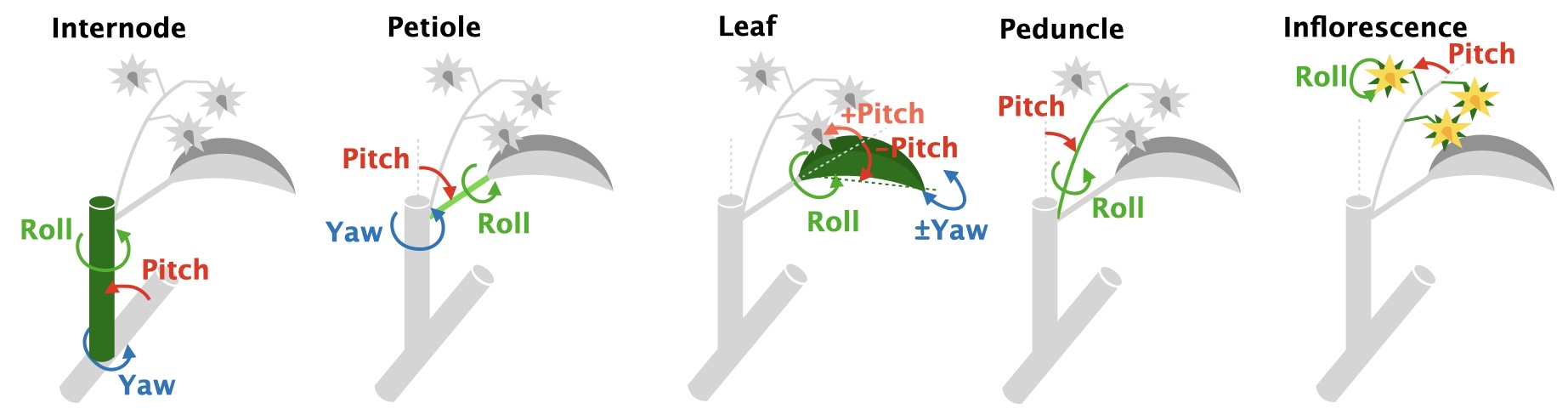}
	\caption{Convention for specifying organ rotations with respect to their parent element.}
	\label{fig:pitchyawrollschematicorgans}
\end{figure}

The orientations of model components are generally specified based on a rotation relative to its parent element. The model uses a pitch-yaw-roll scheme to specify the rotations of elements relative to their parents (Fig. \ref{fig:pitchyawrollschematicorgans}). The meaning of pitch, yaw, and roll angles can vary slightly between organ types, and not all organ types can be rotated based on all three components. In most cases, the default orientation of an organ is such that its axis is aligned with that of its parent organ axis, and the pitch angle specifies a rotation away from the parent axis. The yaw angle specifies a rotation of the child organ about its parent's axis, and the roll angle specifies a rotation about the element's own axis.

\subsection{Organ prototypes}

The morphologies of plant leaves, flowers, and fruit are extremely diverse across the plant kingdom, thus complicating the creation of a universal parameter set to generate all possible geometries. Instead, the approach taken herein is to utilize a generalized framework for creating commonly encountered organ geometries, while also allowing high flexibility to create custom geometries when needed. 

For leaves, there is a built-in routine for procedural leaf prototype generation based on a number of parameters. The routine essentially uses a PNG image with transparency channel to represent the leaf silhouette shape and RGB coloring, and has randomizable parameters controlling other leaf features such as leaf folding, curvature, and waviness, among others. The output is a 3D leaf prototype mesh with user-specified resolution. The advantage of this approach is that it allows for random leaf-to-leaf variation in morphology across the plant, but the disadvantage is that it is not able to represent fine-scale 3D detail in the leaf mesh itself, such as venation patterns. 

When more flexibility is needed, leaf prototype geometries can also be specified by writing an organ `prototype' function that is called each time these organs are generated in the model. The prototype functions should, using any available means, create an instance of the organ of unit size, with its base at the Cartesian origin (0,0,0) and its primary axis aligned with the x-axis. After creating the prototype instance of the organ, the plant architecture model then automatically scales, translates, and rotates the instance as needed to create the appropriate overall plant geometry.

In general, there are two primary means of generating the organ geometry within the prototype functions: 1) manually build the geometry using Helios API calls, or 2) load an existing model from file (usually in Wavefront OBJ format) that was created with third-party software. The advantage of the first approach is that, when integrated with the Helios random number generator, the prototypes can be made procedural with random variation, whereas with the second approach variation between prototypes is more limited and achieved by loading different file versions that vary geometrically. The advantage of the second approach is that sophisticated 3D modeling software (e.g., Blender, Maya, AutoCAD) can be used to create highly detailed and realistic prototype models when such detail is needed. 

Flower and fruit prototypes require specifying a prototype function based on the above criteria, as there is no generic function for generating flower and fruit organs like for leaves. Users can thus generate fruit geometry based on Helios API calls (e.g., adding a spherical object), or by loading a detailed model created in third party software as was mentioned above for leaf models.

\subsection{Phytomer parameters}\label{S:phytomer_parameters}

For each type of phytomer in the model, a set of parameters is defined to determine its structure. The phytomer parameters are distinguished by the fact that they tend to be species- or genotype-specific and are spatially and temporally constant on average within the plant (although they can have random variation). Many parameters have the option of being able to be set as constant values, or set based on a specified distribution (uniform, Gaussian, or Weibull).

The current phytomer parameter set is given in Table \ref{tab:phytomer_parameters}, which is subdivided into parameters of the internode, petiole, leaf, peduncle, and inforescence. Organs generally have rotation parameters, which follow the conventions described above and illustrated in Fig. \ref{fig:pitchyawrollschematicorgans}. As introduced above, leaf, fruit, and flower organs are created by calling prototype functions, which are specified in terms of parameters using a function pointer (which is simply assigned using a string corresponding to the function name). Some more complicated parameter sets are described separately below.

Some parameters are conspicuously missing from the phytomer parameters set, such as the internode radius and length. As mentioned above, phytomer parameters correspond to values that tend to be constant, on average. Values such as the internode radius and length vary with position along the shoot and over time, and are thus considered to be parameters of the shoot not the phytomer. Shoot parameters are described in Sect. \ref{S:shoot_parameters}.

\begin{table}
\caption{Parameter set defining the phytomer.}
\label{tab:phytomer_parameters}
{\scriptsize
\begin{tabularx}{\textwidth}{|c|c|X|}
	\hline
	Parameter & Units & Description \\
	\hline
	\multicolumn{3}{|c|}{internode} \\
	\hline
	\texttt{pitch} & degrees & Angle of the phytomer internode with respect to the previous phytomer along the shoot (creates a zig-zag shoot pattern). \\
	\texttt{phyllotactic\_angle} & degrees & Angle between the petioles/buds of two successive phytomers along the shoot. \\
	\texttt{radius\_initial} & meters & Radius of the internode tube at the time it is created (and will subsequently enlarge based on downstream leaf area). \\
	\texttt{max\_vegetative\_buds\_per\_petiole} & -- & Maximum number of possible vegetative buds per petiole. Some of these buds may not grow depending on the vegetative bud break probability. \\
	\texttt{max\_floral\_buds\_per\_petiole} & - & Maximum number of possible floral buds per petiole. Some of these buds may not grow depending on the flower bud break probability. \\
	\hline
	\multicolumn{3}{|c|}{petiole} \\
	\hline
	\texttt{petioles\_per\_internode} & - & Number of petioles emanating from a single internode (e.g., for an `opposite' growth pattern, = 2). \\
	\texttt{pitch} & degrees & Angle of the petiole base axis with respect to its parent phytomer axis. \\
	\texttt{radius} & meters & Radius of petiole cross-section. \\
	\texttt{length} & meters & Length of the petiole tube. \\
	\texttt{curvature} & degrees/meter & Curvature angle of the petiole per unit length of petiole. If curvature is positive, petiole curves upward toward vertical. If negative, curvature is downward. \\
	\texttt{taper} & - & Ratio between the petiole radius at the tip to the base (e.g., =1 has no taper, =0 comes to a point at the tip). \\
	\hline
	\multicolumn{3}{|c|}{leaf} \\
	\hline
	\texttt{pitch} & degrees & Angle of the leaf axis with respect to its parent petiole axis. \\
	\texttt{yaw} & degrees & Rotation angle of the leaf about its base along the plane of its lamina. \\
	\texttt{roll} & degrees & Rotation angle of the leaf about its own axis (midrib). \\
    \texttt{leaves\_per\_petiole} & - & Number of leaves per petiole. $>$1 creates a compound leaf. \\
\texttt{	leaflet\_offset} & - & If a compound leaf (leaves\_per\_petiole$>$1), this sets the spacing between adjacent leaflets along the petiole as a fraction of the petiole length. \\
	\texttt{leaflet\_scale} & - & If a compound leaf (leaves\_per\_petiole$>$1), this sets the scaling factor of the leaflet moving down the petiole with respect to the previous leaf ($<$1 scales down, $>$1 scales up).\\
	\texttt{prototype\_scale} & - & Scaling factor applied to the leaf prototype. Usually the prototype has unit length, so this sets the physical length of the leaf. \\
	\texttt{prototype}  & \texttt{LeafPrototype} struct & Structure containing information to build leaf prototypes. \\
	\hline
		\multicolumn{3}{|c|}{peduncle} \\
		\hline
	    \texttt{length} & meters & Length of the peduncle (inflorescence supporting structure). \\
		\texttt{radius} & meters & Radius of the peduncle. \\
		\texttt{pitch} & degrees & Angle of the peduncle axis with respect to its parent internode axis. \\
		\texttt{roll} & degrees & Rotation angle of the peduncle about its own axis. \\
		\texttt{curvature} & degrees/meter & Curvature angle of the peduncle per unit length of peduncle. \\
		\hline
		\multicolumn{3}{|c|}{inflorescence} \\
		\hline
		\texttt{pitch} & degrees & Angle of the fruit axis relative to its parent peduncle axis. \\
		\texttt{roll} & degrees & Rotation angle of the fruit about its own axis (x-axis of fruit prototype). \\
	    \texttt{flowers\_per\_peduncle} & - & Number of flowers per peduncle (rachis). \\
		\texttt{flower\_offset}  & - & If peduncle has multiple flowers/fruit (flowers\_per\_peduncle$>$1), this sets the spacing between adjacent flowers/fruit along the peduncle as a fraction of the peduncle length. \\
		\texttt{flower\_prototype\_scale} & - & Scaling factor applied to the flower prototype. Usually the prototype has unit length, so this sets the physical length of the flower. \\
		\texttt{flower\_prototype\_function} & - & Pointer to a function that generates the flower prototype. \\
		\texttt{fruit\_prototype\_scale} & - & Scaling factor applied to the fruit prototype. Usually the prototype has unit length, so this sets the physical length of the fruit. \\
		\texttt{fruit\_prototype\_function} & - & Pointer to a function that generates the fruit prototype. \\
		\hline
	\end{tabularx}
}
\end{table}

\subsubsection{Phyllotaxis}

\begin{figure}
	\centering
	\includegraphics[width=\textwidth]{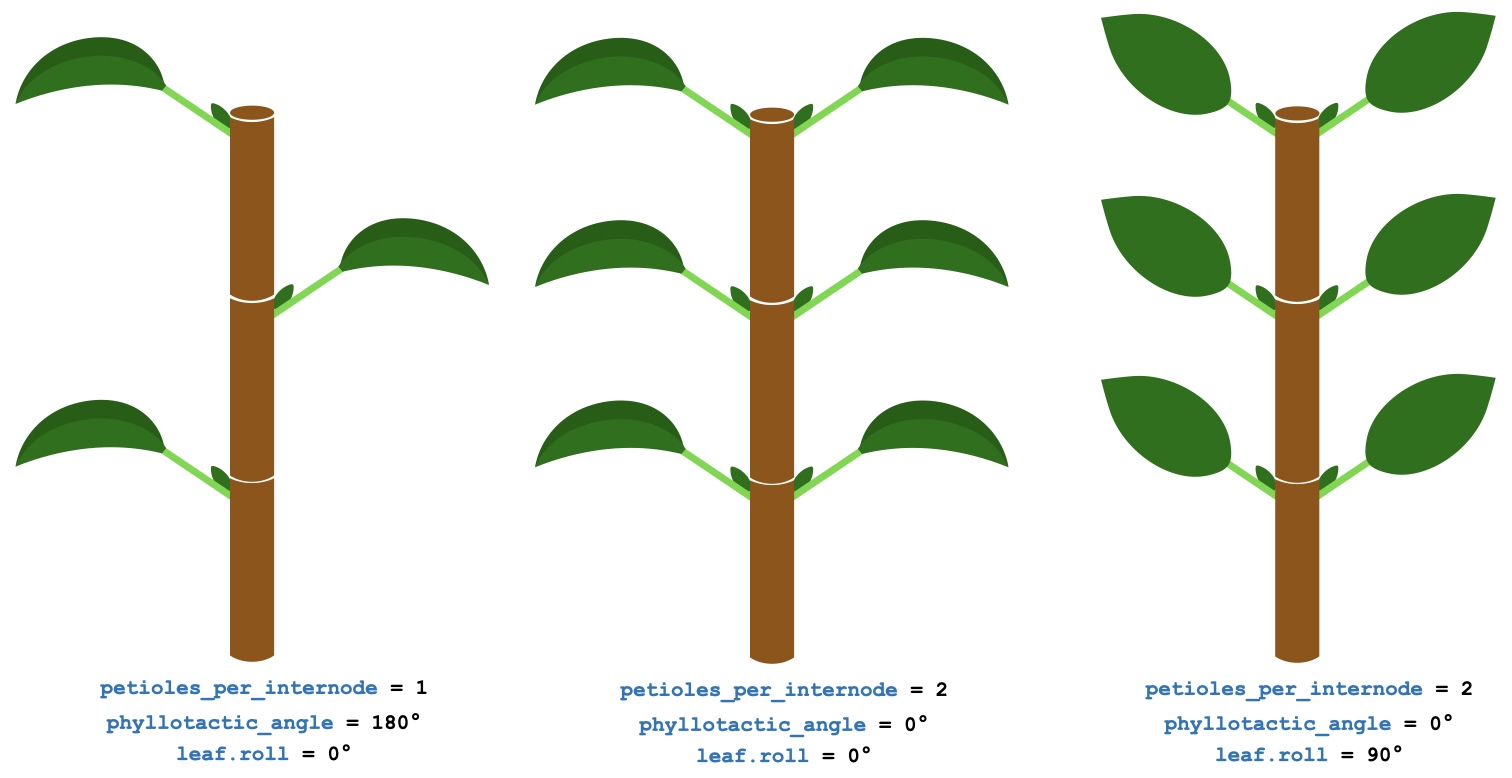}
	\caption{Schematic illustration of phytomer parameter combinations leading to different phyllotaxis behavior. In this example, the parameters \texttt{internode.phyllotactic\_angle}, \texttt{petiole.petioles\_per\_internode}, and \texttt{leaf.roll} are modified to change the overall phyllotaxis.}
	\label{fig:phyllotaxis}
\end{figure}

Phyllotaxis is the arrangement of leaves along their parent stem, which is determined in the model through two primary phytomer parameters: \texttt{internode.phyllotactic\_angle} and \texttt{petiole.petioles\_per\_internode} (Fig. \ref{fig:phyllotaxis}). The phyllotactic angle specifies the roll rotational angle of the phytomer relative to the previous phytomer along the shoot. An "alternate" growth pattern is achieved by setting \texttt{petiole.petioles\_per\_internode} to 1 and \texttt{internode.phyllotactic\_angle} to 180$^\circ$ (Fig. \ref{fig:phyllotaxis}a), which is common in many vegetative crops like tomato and legumes. An "opposite" growth pattern is achieved by setting \texttt{petiole.petioles\_per\_internode} to 2 and \texttt{internode.phyllotactic\_angle} to 0 (Fig. \ref{fig:phyllotaxis}b), which occurs in some tree species like olive and in herbs like basil and mint. Other growth patterns are achievable by adjusting these parameters, for example a spiral growth pattern by setting \texttt{internode.phyllotactic\_angle} to an intermediate angle that is not a multiple of 180$^\circ$ (as is common in flowers like sunflower or trees like almond).

\subsubsection{Compound leaves}

\begin{figure}
	\centering
	\includegraphics[width=\textwidth]{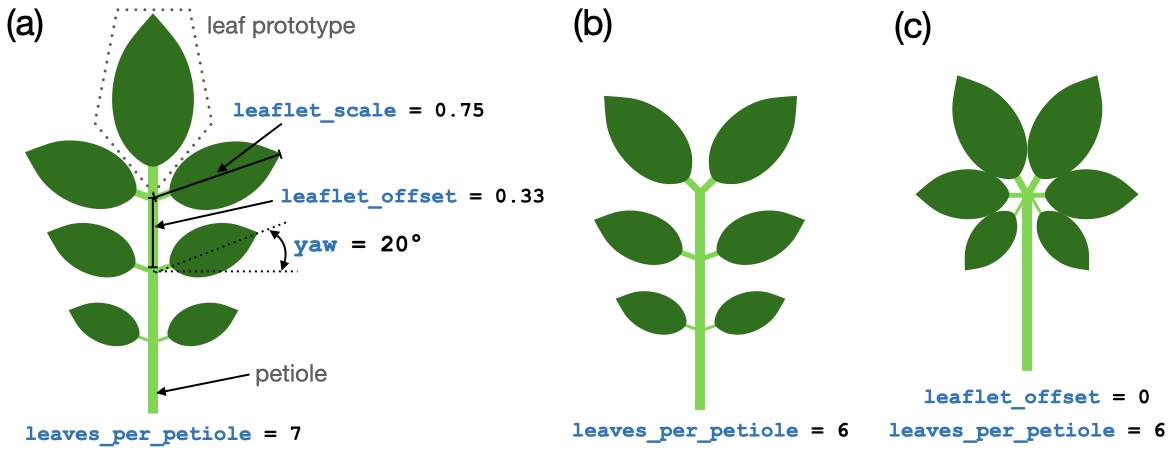}
	\caption{Schematic illustration of phytomer parameters determining the structure of compound leaves. }
	\label{fig:compound_leaves}
\end{figure}

Many species have compound leaves, characterized by multiple leaflets per petiole (Fig. \ref{fig:compound_leaves}). The model has two primary means of creating a compound leaf. The first would be to have the leaf prototype function create the entire compound leaf together in a single function call. This can allow for precise control of the appearance of the compound leaf, but with the downside that it does not allow for stochastic variation within the leaflet itself. For this approach, the user would set \texttt{leaf.leaves\_per\_petiole} to 1, and one leaf prototype (which contains the whole compound leaf geometry) would be placed at the end of the petiole. 

The other method, which does allow for random variability, is to set \texttt{leaf.leaves\_per\_petiole} $>1$ along with other phytomer parameters that control the compound leaf geometry. In this case, the leaf prototype function is called independently for each leaflet along the petiole, which means that any random leaf parameters like the leaf scale, rotations, etc. are sampled for each leaflet. If \texttt{leaf.leaves\_per\_petiole}$>1$, additional parameters of \texttt{leaf.leaflet\_offset} and \texttt{leaf.leaflet\_scale} are considered. \texttt{leaf.leaflet\_offset} determines the spacing between successive pairs of leaflets as a fraction of the overall petiole length. \texttt{leaf.leaflet\_scale} applies a scaling factor to each pair of leaflets moving down the petiole as a fraction of the size of the previous leaflet moving from the tip. The tip leaf is not additionally scaled, and it ignores any leaf rotations. 

For some species, the morphology of leaflets changes substantially along the petiole, such as in tomato or legumes for example. If using the first method described above, this is not an issue because the user can simply create the entire compound leaf structure manually such that this variation is included. For the second method, there is an argument to the leaf prototype function that specifies the index of the leaflet from the tip, such that different behavior can be applied based on this index value such as using a different image texture or loading a different polygon model from file. Examples of this can be found in the leaf prototype functions for common bean and cowpea plant models in the default species library.

\subsubsection{Compound inflorescence}

Analogous to leaflets along the petiole of a compound leaf, the phytomer's reproductive structures can potentially have multiple flowers/fruit along the peduncle. As with compound leaves, the inflorescence structure can either be created manually (with static variation), or procedurally based on parameters in the model. To use the former approach, the user would set \texttt{inflorescence.flowers\_per\_peduncle} to 1 and define the entire inflorescence structure in a single call to the flower or fruit prototype function. In the latter approach, \texttt{inflorescence.flowers\_per\_peduncle} is set equal to the potential (maximum) number of flower buds per peduncle. The phyllotaxis of the floral buds follows that of the parent phytomer it originates from. The parameter \texttt{inflorescence.flower\_offset} defines the spacing of successive flower buds along the peduncle, where the value gives the spacing as a fraction of the total peduncle length.

\subsubsection{Custom phytomer creation and callback functions}

In some cases, the usual phytomer behavior may need to be overridden to represent certain plant species. To achieve this, users can define a custom phytomer creation function that is called at the time the phytomer is created. An example where this might be used is in vegetative plants to modify the number of axillary vegetative buds as a function of phytomer position along the shoot, or to remove buds along the shoot (i.e., creating "blind" nodes).

There is also an additional custom callback function that can optionally be defined, which is called for each phytomer at every time step. One example of where this is needed is in creating the model of redbud trees {\em Cercis canadensis}, where after dormancy the first several axillary buds from the tip are all vegetative, then the rest are floral. An additional complication is that the location of which buds should be floral or vegetative is not known when the phytomer is created, but only after the plant has entered dormancy. The phytomer callback function for this species checks whether the phytomer has entered dormancy, and if so it enforces this bud pattern by removing the first several axillary floral buds from the tip, then removing the rest of the vegetative buds for all other axillary nodes.

\subsubsection{Visualization/rendering parameters}

Additional phytomer parameters are available to specify the visual appearance of phytomer elements, which are not listed in Table \ref{tab:phytomer_parameters} for brevity, but can be found in the Helios documentation. These include parameters for specifying the RGB color or texture map image for determining the color of geometric elements, and parameters to control the subdivision resolution of organs. The resolution parameters are important for controlling the total primitive count of the model. Increasing the resolution will generally make the model geometry look smoother, but will increase the memory and computational requirements for the model. Note that in some cases the resolution parameter is not used for certain organs, for example if the organ prototype function loads a static 3D model from file, its resolution is not adjustable.

\subsection{Shoot parameters}\label{S:shoot_parameters}

A shoot consists of one or more phytomers connected end-to-end in series (Fig. \ref{fig:phytomerschematic}). The shoot can evolve dynamically via elongation or girth increase of the internode, or by spawning new phytomers from vegetative buds located at the shoot tip or axillary to petioles. Characteristics of the shoot may also vary spatially across the plant, such as in acrotonic growth (see below). This spatiotemporal variation in parameters distinguishes shoot parameters from phytomer parameters. An exhaustive list of shoot parameters is given in Table \ref{tab:shoot_parameters}.

\begin{table}
	\caption{Parameter set defining the shoot.}
	\label{tab:shoot_parameters}
	{\scriptsize
		\begin{tabularx}{\textwidth}{|c|c|X|}
			\hline
			Parameter & Units & Description \\
			\hline
			\multicolumn{3}{|c|}{geometric parameters} \\
			\hline
			%\texttt{phytomer\_parameters} & - & Parameters defining the geometry of the phytomers comprising this shoot. \\
			\texttt{max\_nodes} & - & Maximum number of nodes/phytomers along a shoot. \\
			\texttt{max\_nodes\_per\_season} & - & Maximum number of nodes/phytomers produced by a shoot in a single season ($\leq$\texttt{max\_nodes}). \\
			\texttt{internode\_radius\_initial} & meters & Initial radius of the internode when created. \\
			\texttt{insertion\_angle\_tip} & degrees & Angle of the child shoot with respect to the parent shoot at the tip of the parent shoot. \\
			\texttt{insertion\_angle\_decay\_rate} & degrees/node & Rate of increase of the child insertion angle moving down the parent shoot. \\
			\texttt{internode\_length\_max} & meters & Maximum length (with respect to position along the parent shoot) of the internode of a child shoot. \\
			\texttt{internode\_length\_min} & meters & Minimum length (with respect to position along the parent shoot) of the internode of a child shoot. \\
			\texttt{internode\_length\_decay\_rate} & meters/node & Rate of decrease of the internode length moving down the parent shoot. \\
			\texttt{base\_roll} & degrees & Roll angle of the shoot, which effectively specifies the angle of the first petiole relative to the parent shoot. \\
			\texttt{base\_yaw} & degrees & Shoot yaw angle relative to the parent shoot. \\
			\texttt{gravitropic\_curvature} & degrees/meter & Curvature angle of the shoot per unit length of shoot. If curvature is positive, shoot curves upward toward vertical. If negative, curvature is downward. \\
			\texttt{tortuosity} & degrees / (meters)$^{1/2}$ & Factor determining the amount of random "wiggle" in internode growth along the shoot. \\
			\hline
			\multicolumn{3}{|c|}{growth parameters} \\
			\hline
			\texttt{phyllochron\_min} & days/leaf & Time between the emergence of successive phytomers along the shoot. \\
			\texttt{elongation\_rate\_max} & meter/meter/day & Relative rate of elongation of the internode of the shoot. Units are meters of elongation per meter of maximum internode length per day. \\
			\texttt{girth\_area\_factor} & cm$^2$ / m$^2$ & Cross-sectional area of internode (girth), determined by the amount of downstream leaf area. The girth will only increase and does not decrease if leaves are lost. Set this factor to 0 to prevent girth scaling. \\
			\texttt{vegetative\_bud\_break\_time} & days & Amount of time after bud is created or after dormancy is broken for vegetative bud break. \\
			\texttt{vegetative\_bud\_break\_probability\_decay\_rate} & 1/nodes & Rate of reduction/increase along the shoot in probability of a vegetative bud breaking dormancy and emerging as a shoot. \\
			\texttt{vegetative\_bud\_break\_probability\_min} & - & Minimum probability along the shoot of a bud breaking dormancy and emerging as a shoot. \\
			\texttt{max\_terminal\_floral\_buds} & - & Potential number of terminal floral buds on a shoot. \\
			\texttt{flower\_bud\_break\_probability} & - & Probability of a flower bud emerging as a flower. \\
			\texttt{fruit\_set\_probability} & - & Probability of a flower setting into a fruit. \\
			\texttt{growth\_requires\_dormancy} & - & Whether vegetative buds require winter dormancy before shoot emergence. If true, emergence follows dormancy; if false, it may occur in the same season. \\
			\texttt{flowers\_require\_dormancy} & - & Whether flower buds require winter dormancy before emergence. If true, emergence follows dormancy; if false, it may occur in the same season. \\
			\texttt{determinate\_shoot\_growth} & - & Whether shoot growth is determinate. If true, growth stops at flowering and the apical bud becomes dormant; if false, growth continues after flowering. \\
				\hline
			\end{tabularx}
		}
	\end{table}

\subsubsection{Shoot geometric parameters}

The shoot parameters structure stores a copy of a phytomer parameters structure, which defines the phytomer "template" it will use when creating the shoot structure. The plant may consist of multiple different shoot types, each potentially utilizing a different set of phytomer parameters. The shoot structure is created by replicating phytomers based on that shoot type's phytomer parameters. 

\textit{Node count:} The absolute maximum number of nodes/phytomers a shoot can have is limited by the parameter \texttt{max\_nodes}. The maximum number of nodes that a shoot can add in a given season is limited by the parameter \texttt{max\_nodes\_per\_season}. This allows perennial shoot growth to cease at some specified point, but continue growing the following season.

\textit{Child shoot insertion angle:} The angle of a child shoot relative to its parent shoot is the insertion angle (= 0 if parallel). The insertion angle can be specified as a function of position along the shoot, or held constant. The parameter \texttt{insertion\_angle\_tip} specifies the insertion angle at the tip of the shoot. The tip of the shoot is defined as the most distal phytomer on the shoot at the time the vegetative bud for which we are calculating insertion angle breaks. The parameter \texttt{insertion\_angle\_decay\_rate} determines the rate of increase in the insertion angle along the shoot, in units of degrees per node. The insertion angle can increase at this rate until it reaches a value of 90 degrees.

\textit{Internode length:} The maximum length that an internode can achieve when fully-elongated is specified by three parameters that control how the length changes along the shoot. The fully-elongated internode length can either increase (acrotonic growth) or decrease (basitonic growth) along the shoot. If \texttt{internode\_length\_decay\_rate} $<$ 0, this corresponds to basitonic growth and the fully-elongated internode length will start at a value of \texttt{internode\_length\_max} at the shoot base and decrease at a rate of \texttt{internode\_length\_decay\_rate} meters/node until reaching a minimum value of \texttt{internode\_length\_min} that persists for the further length of the shoot. Setting \texttt{internode\_length\_decay\_rate} $>$ 0 gives acrotonic growth with the maximum internode length occurring at the shoot tip (also defined at the time that the vegetative bud breaks). Setting \texttt{internode\_length\_decay\_rate} $=$ 0 results in all fully-elongated internode lengths equal to a constant value of \texttt{internode\_length\_max}.

\textit{Rotation:} The base rotation of the shoot can be set with the \texttt{base\_roll} and \texttt{base\_yaw} parameters. The yaw rotates the child shoot about its parent shoot's axis, and the roll rotates the child shoot about its own axis (default is that the shoot roll is such that the first petiole along the shoot points in the direction of its parent shoots' axis).

\textit{Curvature:} The parameter \texttt{gravitropic\_curvature} determines the propensity of the shoot to curve toward the vertical axis as it grows. The curvature is applied to an internode at the time it is created by modifying the zenith angle of the new phytomer internode with respect to its parent phytomer internode. This curvature adjustment angle of the internode is given by

\begin{equation}
	\Delta \theta_z = s\left(1-\mathrm{cos}\,\theta_z \right)L_i \times (\texttt{gravitropic\_curvature}),
\end{equation}

\noindent where $\theta_z$ is the angle between the internode axis at the shoot tip and the vertical axis, $L_i$ is the fully-elongated internode length at the shoot tip, and $s=1$ if the shoot tip is pointing downward (i.e., $\theta_z>90^\circ$), and $s=0.5$ otherwise. This produces the effect that the shoot will curve strongest when the shoot is pointing downward, and the rate of curvature will gradually decline as the shoot tip approaches vertical.

\textit{Random walk:} The model uses an analogue to Brownian motion (random walk) to give stochastic variation in the shoot path. As with shoot curvature, the stochastic variation is applied at the time the phytomer is created by modifying the angle of the new phytomer internode with respect to its parent phytomer internode.  The random walk is applied to the internode's zenith angle ($\theta_z$; angle between internode axis and vertical) and to the internode's yaw ($\theta_y$; rotation angle about the vertical axis) according to

\begin{align}
	 \theta_z^{t+1} &= \theta_z^t - \frac{1}{2}\theta_z^t L_i + (\texttt{tortuosity})\times dW_z, \\
	 \theta_y^{t+1} &= \theta_y^t - \frac{1}{2}\theta_y^t L_i + (\texttt{tortuosity})\times dW_y, 
\end{align}

\noindent where superscript $t$ denotes the angle at the previous phytomer internode, and superscript $t+1$ denotes the angle for the new phytomer internode, $dW$ is a random number drawn from Gaussian distribution with mean of zero and variance of $L_i$ (the fully-elongated internode length), and \texttt{tortuosity} is the shoot parameter used to increase stochastic variation. In essence, the term containing the \texttt{tortuosity} parameter adds stochastic variation, and the middle term dampens out this variation.

\subsubsection{Growth parameters}

\textit{Growth via new phytomer production: } New phytomers are produced from an active terminal vegetative bud at a rate of 1/\texttt{phyllochron\_min}, which is the primary means by which the plant "grows". 

\textit{Growth via elongation:} When a new phytomer is produced, the maximal length of the internode and leaf/petiole is pre-determined. The length of a new phytomer internode is initialized at 1\% of its final, fully-elongated length. Similarly, the petiole and leaf are initially scaled to be 1\% of their fully-elongated length. Over time, these organs elongate at a rate determined by the \texttt{elongation\_rate\_max} parameter, which specifies the maximum relative elongation rate in units of meters per meter of maximum length per day (m$\cdot$m$^{-1}\cdot$day$^{-1}$). Elongation continues until the organ reaches its pre-determined maximum length. 

\textit{Radial internode growth:} When a new phytomer is created, it starts with an internode radius of \texttt{internode\_radius\_initial}. This internode radius increases over time according to the amount of downstream leaf area that is created (downstream meaning any future leaf area created distally along its shoot, or on any future child shoots). The ratio of internode cross-sectional area to downstream leaf area is given by the parameter \texttt{girth\_area\_factor} (cm$^2$ cross-sectional area per m$^2$ downstream leaf area). If \texttt{girth\_area\_factor} is set to 0, the internode radius will be locked at \texttt{internode\_radius\_initial}.

\textit{Growth via new shoots}: When a vegetative bud breaks, a new phytomer is created which forms the basis of a new shoot. The probability that a given axillary vegetative bud breaks and leads to a new shoot is given by the parameter \texttt{vegetative\_bud\_break\_probability} (note that apical vegetative buds do not adhere to this probability value). In annual plants, vegetative buds break in the same season in which they were created. For perennial plants, vegetative buds may break in the same season in which they were created (e.g., sylleptic shoots), or they may require a dormancy period in order to produce a shoot (e.g., proleptic shoots). Some perennial plants may have both types of shoots. This behavior is controlled by the \texttt{growth\_requires\_dormancy} parameter (boolean), which is set to true if the shoot type's buds require a winter dormancy period to produce a shoot, in which case the bud will break after a time period of \texttt{bud\_break\_time} after dormancy. If this parameter is set to false, vegetative buds will break \texttt{bud\_break\_time} after the bud is created.

Apical buds of indeterminate shoots (\texttt{determinate\_shoot\_growth} set to false) will continue to generate new phytomers until the shoot phytomer count reaches \texttt{max\_nodes} (or the apical bud is manually killed). Determinate shoots (\texttt{determinate\_shoot\_growth} set to true) will continue growing until either the max nodes is reached, or the shoot begins flowering (see Sect. \ref{S:phenology} below for control of flowering).

\textit{Reproductive growth}: The parameter \texttt{flower\_bud\_break\_probability} determines the probability that a floral bud will form a flower. Similar to vegetative buds, the parameter \texttt{flowers\_require\_dormancy} determines whether the floral bud will break in the same season it is created, or if a winter dormancy is required. The parameter \texttt{fruit\_set\_probability} determines the probability a flower will set into a fruit.

\subsection{Phenological transitions}\label{S:phenology}

As the plant grows, it may transition to different stages of its growth cycle (i.e., phenology). There are 6 parameters that can be set to control the time to various phenological transitions. Setting any of these parameters to a negative number will disable that particular transition.

When the plant is created, it is in a dormant state by default unless overridden. The parameter \texttt{time\_to\_dormancy\_break} determines the amount of time before dormancy is broken and new leaves may start to appear. For perennial plants, this parameter also applies to growth in future seasons - it specifies the length of the winter dormancy period. 

If flowering is enabled, closed flowers will appear after \texttt{time\_to\_flower\_initiation} after dormancy is broken. These flowers will open after a time of \texttt{time\_to\_flower\_opening} following flower initiation. If \texttt{time\_to\_flower\_initiation} is set to a negative value (e.g., because the user does not want to explicitly represent closed flowers), then open flowers will appear after a time of \texttt{time\_to\_flower\_opening} following dormancy break. 

Flowers will turn into fruit after a time of \texttt{time\_to\_fruit\_set} following the time of previous phenological transition (dormancy break, flower initiation, or flower opening, depending on what has been enabled by the user). After appearing, fruit continuously grows over time, and eventually reaches its full size after a time of \texttt{time\_to\_fruit\_maturity}. 

The growing season ends after a period of \texttt{time\_to\_senescence} after dormancy break, and the plant will enter dormancy. For annual or deciduous plants, this means that leaves will be removed. There is an additional flag that can be enabled to create an evergreen plant, which will not drop its leaves when it enters dormancy but will stop any further growth.

\subsection{Collision avoidance}

The plant architecture model integrates the Helios collision detection plug-in in order to enable two primary types of object collision avoidance: 1) "soft" avoidance of other plant organs during growth in order to create a plant structure that generally minimizes collisions with itself and other plants while also tending to fill open space (i.e., spatial colonization), and 2) "hard" avoidance of solid obstacle objects that should not be intersected like the ground or buildings.

\begin{figure}
	\centering
	\includegraphics[width=0.6\textwidth]{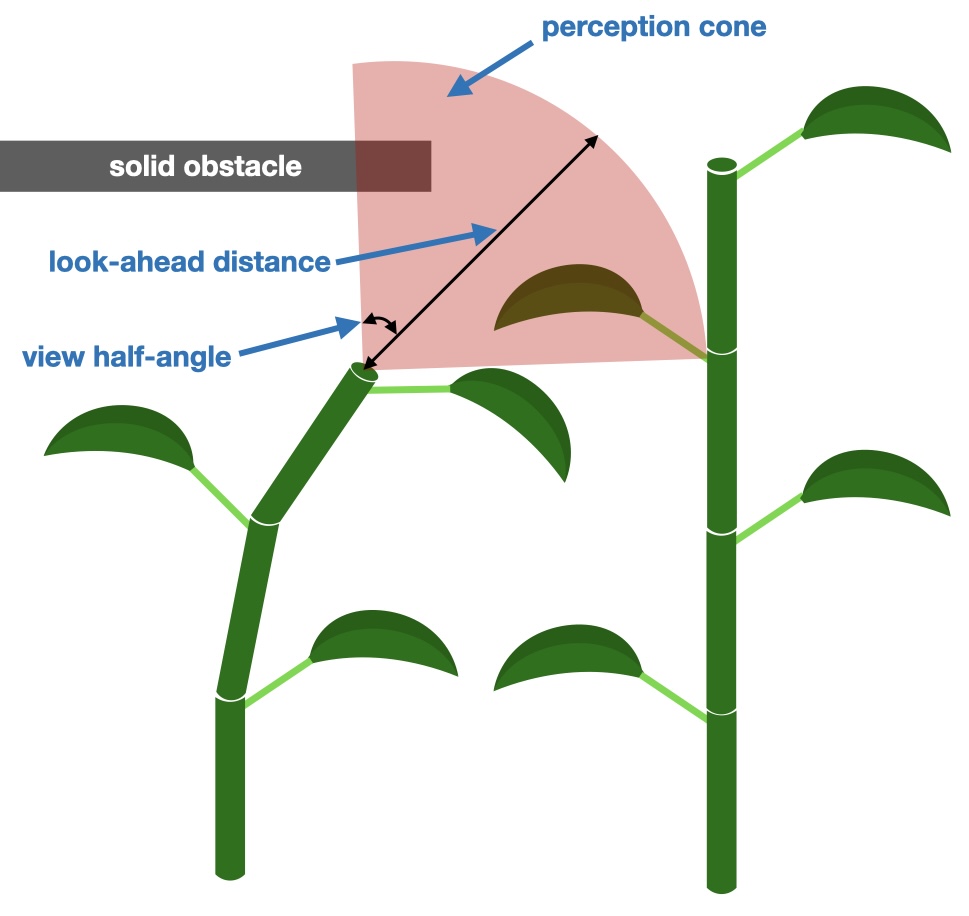}
	\caption{Schematic representation of the perception cone used to determine potential collisions at the shoot apex in order to modify the internode and/or petiole growth direction to minimize or eliminate collisions.}
	\label{fig:collision_perception_cone}
\end{figure}

Both types of collision avoidance use the concept of a perception cone at the shoot apex (Fig. \ref{fig:collision_perception_cone}). The parameters of the cone are user-defined, which determine the "look-ahead distance" (length of cone axis) and the cone field of view. Ray-tracing calculations are used to determine objects `sensed' within the cone view. It should be noted that the parameters of the perception cone affects model computational expense. A larger cone will tend to result in reduced performance because more intersection queries will be needed in general. If the cone is too small, model accuracy may be sacrificed because the plant may not adequately perceive obstacles and thus growth may not be appropriately modified.

The collision detection framework utilizes several optimizations to improve performance. It uses OpenMP parallelization to accelerate ray-tracing operations associated with perception cone object detection. It also uses an efficient plant-centric bounding volume hierarchy (BVH) to rapidly cull distant geometry during ray intersection traversals.

\begin{figure}
	\centering
	\includegraphics[width=\textwidth]{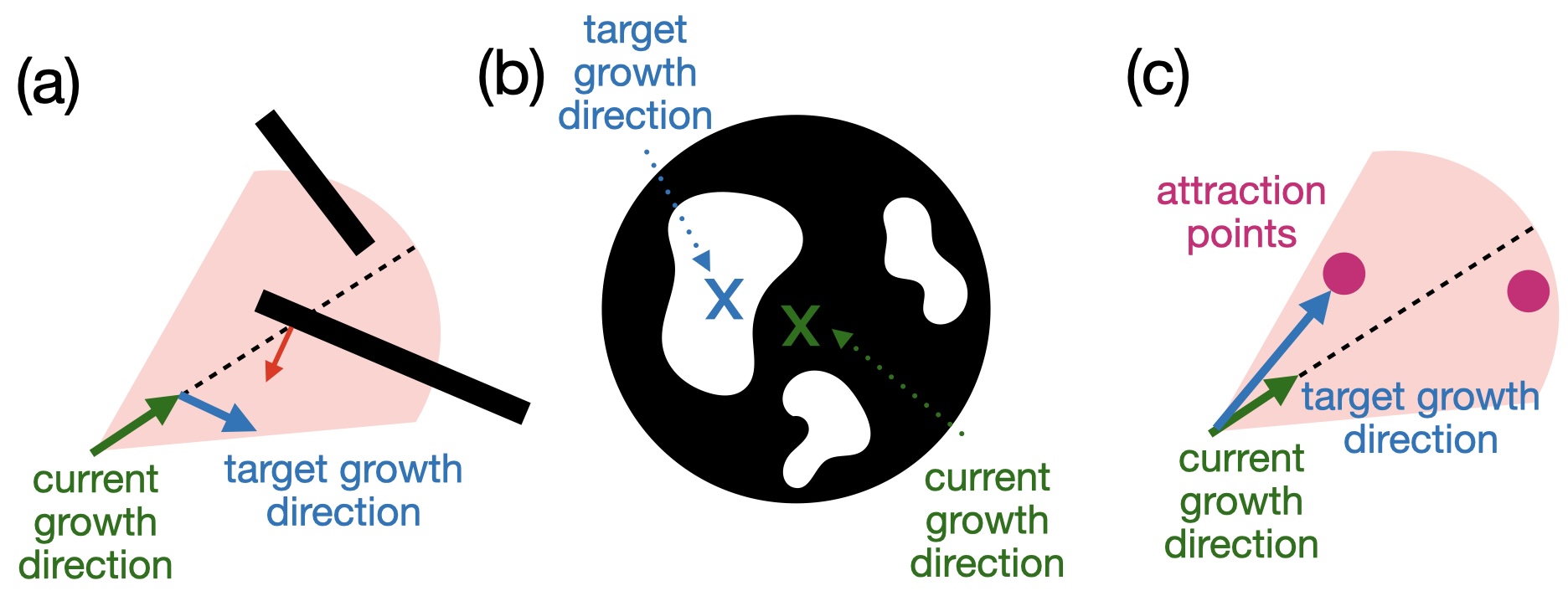}
	\caption{Schematic representation of how the perception cone is utilized to determine the augmented (target) growth direction. (a) "hard" collision avoidance: perception cone (pink) points in the initial direction of internode, petiole, or peduncle growth. The normal vector of the nearest object detected in the cone is used to calculate a target growth direction perpendicular to the object normal. (b) "soft" collision avoidance: hemispherical projection of the perception cone field of view. Target growth direction is determined based on a weighted combination of the perceived gap size and distance from the cone axis. (c) attraction points: attraction points lying in the perception cone are detected. The target growth direction points toward the closest detected attraction point.}
	\label{fig:collision_types}
\end{figure}

1. \underline{Soft growth collision avoidance}: Calling \texttt{enableSoftCollisionAvoidance()} will enable "soft" collision avoidance for internodes, and optionally for petioles and peduncles. The term "soft" here is used because growth will generally tend in a direction that minimizes collisions, but it will not strictly prevent collisions. This avoids a costly iterative procedure that would be needed to globally ensure there are no collisions.

When enabled, each time a new internode (or optionally petiole or peduncle) is spawned during growth, a "perception cone" with axis oriented in the initial candidate growth direction is used to determine potential obstacles.  A user-defined number of rays are launched from the cone apex toward its base to calculate points of intersection for objects within the cone. The algorithm then uses this information to determine the largest contiguous gap within the cone's view, and a direction vector originating at the cone's base and pointing to the middle of the gap. An "inertia" factor is set that determines how abruptly growth is adjusted toward the  gap -- a factor of 0.0 sets the new growth direction exactly toward the gap, while a factor of 1.0 does not modify the growth direction. By default, only leaves are considered as obstacles for soft collision avoidance, but other organ types can be optionally enabled, which are accompanied by increased computational cost. 

2. \underline{Hard boundary collision avoidance}: To strictly enforce that plants do not grow into solid boundaries, "hard" boundary collision avoidance can be enabled using \texttt{enableSolidObstacleAvoidance()}. Users define which geometry should be considered as solid boundaries, and the parameters defining the perception cone. Similar to soft object avoidance, solid obstacle avoidance involves launching rays from the perception cone apex toward its base to determine solid obstacles contained in the perception cone. If present, the distance to the solid obstacle closest to the cone apex is determined, along with the normal vector of the obstacle. The new target growth direction is a vector perpendicular to the surface of the object. The strength of the change in growth direction to achieve the target growth direction increases as the plant gets closer to the object. When it gets very near the object, the growth direction is set equal to the target growth direction to ensure that the object will not be hit.

When soft collision avoidance is enabled (1 above), depending on the value of the inertia factor, plants may effectively avoid solid boundaries without hard boundary collision avoidance enabled. 

When enabling hard boundary collision avoidance, users can optionally enable fruit adjustment based on solid obstacle collisions. This is useful for large fruit growing near solid obstacles (e.g., large fruit resting on the ground). When this is enabled, fruit are rotated away from the boundary such that a bounding box encompassing the fruit no longer intersects the solid boundaries it was intersecting. Then an iterative refinement procedure rotates the fruit back toward the boundary to get it closer without intersecting it again. 

\subsection{Attraction points}

Plant growth can be constrained using arbitrarily defined "attraction points" (Fig. \ref{fig:collision_types}c). Attraction points are specified as a list of Cartesian points in space. As growth proceeds, a perception cone originating from growing shoot tips is used to query attraction points existing in the cone. If an attraction point is found in the cone, the target growth direction becomes a vector originating at the cone apex and pointed toward the nearest attraction point in the cone. Like the "soft" object avoidance described above, a user-defined inertia factor determines how abruptly growth will change in order to match the target growth direction. 

Some plant models in the existing plant library (Sect. \ref{sect:plant_library}) utilize attraction points by default to constrain their growth. In this case, each plant model has its own instance of attraction points and parameters that are kept separate from the main interface exposed to the user. This allows users to define additional attraction points and associated parameters without affecting library plant builds.

\section{Model usage}

This section presents a brief overview of how the plant architecture plug-in is actually used to build dynamic plant models. These examples are by no means exhaustive, and are not meant to be a full documentation of the model. The goal is simply to give examples that illustrate the various ways in which plants can be built from the plug-in.

\subsection{Loading a plant from the existing library}\label{sect:plant_library}

At the time of this writing, Helios plant architecture model has 26 different plant models in the existing library that can be readily loaded, which spans various tree, annual, and weed plant species. Table \ref{tab:plant_library} summarizes the existing plant library. Detailed renderings of example plant models are shown in Figs. \ref{fig:bean_growth}-\ref{fig:manyplants}, and in the Appendix Figs. \ref{fig:example_visualization_apple_house}-\ref{fig:example_visualization_bougainvillea}.

\begin{table}[htbp]
	\centering
	\caption{Plant types and their corresponding string identifiers}
	\begin{tabular}{|l|l|}
		\hline
		\textbf{Plant type/variety} & \textbf{Plant type string argument} \\
		\hline
		Almond Tree (\textit{Prunus dulcis}) & almond \\ \hline
		Apple Tree (\textit{Malus pumila}) & apple \\ \hline
		Bindweed (\textit{Convolvulus arvensis}) & bindweed \\ \hline
		Bougainvillea (\textit{Bougainvillea spectabilis}) & bougainvillea \\ \hline
		Butter Lettuce (\textit{Lactuca sativa}) & butterlettuce \\ \hline
		Capsicum Pepper (\textit{Capsicum annuum}) & capsicum \\ \hline
		Cheeseweed (\textit{Malva neglecta}) & cheeseweed \\ \hline
		Common Bean (\textit{Phaseolus vulgaris}) & bean \\ \hline
		Cowpea (\textit{Vigna unguiculata}) & cowpea \\ \hline
		Eastern Redbud (\textit{Cercis canadensis}) & easternredbud \\ \hline
		Grapevine (\textit{Vitis vinifera}; VSP trellis) & grapevine\_VSP \\ \hline
		Grapevine (\textit{Vitis vinifera}; Wye trellis) & grapevine\_Wye \\ \hline
		Ground Cherry (\textit{Physalis philadelphica}) & groundcherryweed \\ \hline
		Maize (\textit{Zea mays}) & maize \\ \hline
		Olive Tree (\textit{Olea europaea}) & olive \\ \hline
		Pistachio Tree (\textit{Pistachia vera}) & pistachio \\ \hline
		Puncturevine (\textit{Tribulus terrestris}) & puncturevine \\ \hline
		Rice (\textit{Oryza sativa}) & rice \\ \hline
		Sorghum (\textit{Sorghum bicolor}) & sorghum \\ \hline
		Soybean (\textit{Glycine max}) & soybean \\ \hline
		Strawberry (\textit{Fragaria \(\times\) ananassa}) & strawberry \\ \hline
		Sugar Beet (\textit{Beta vulgaris}) & sugarbeet \\ \hline
		Tomato (\textit{Solanum lycopersicum}) & tomato \\ \hline
		Cherry Tomato (\textit{Solanum lycopersicum}) & cherrytomato \\ \hline
		Walnut Tree (\textit{Juglans regia}) & walnut \\ \hline
		Wheat (\textit{Triticum aestivum}) & wheat \\ \hline
	\end{tabular}
	\label{tab:plant_library}
\end{table}

\begin{figure}
	\centering
	\includegraphics[width=0.75\textwidth,trim={0cm 0 6cm 6cm},clip]{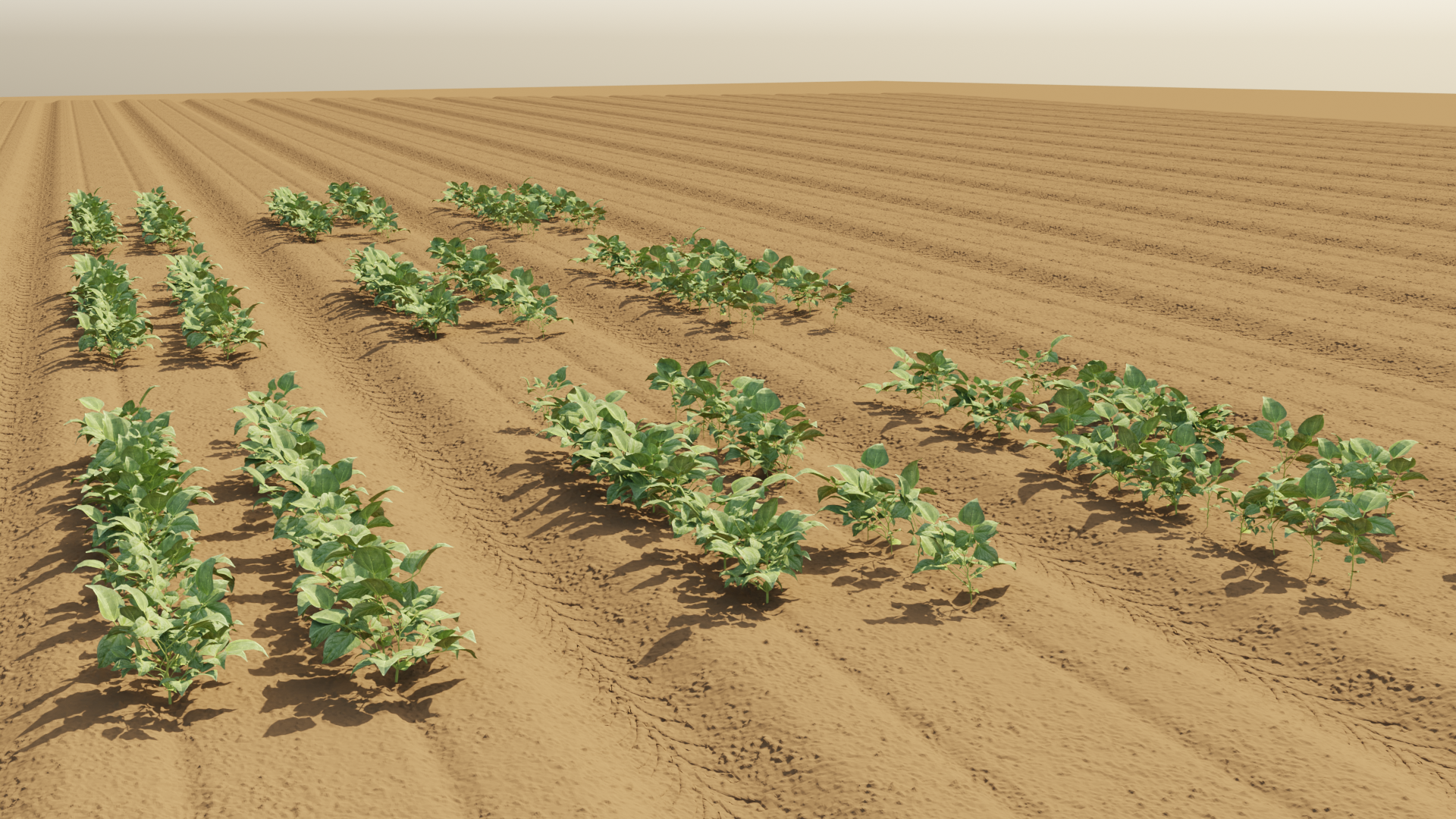}
	\includegraphics[width=0.75\textwidth,trim={0cm 0 6cm 6cm},clip]{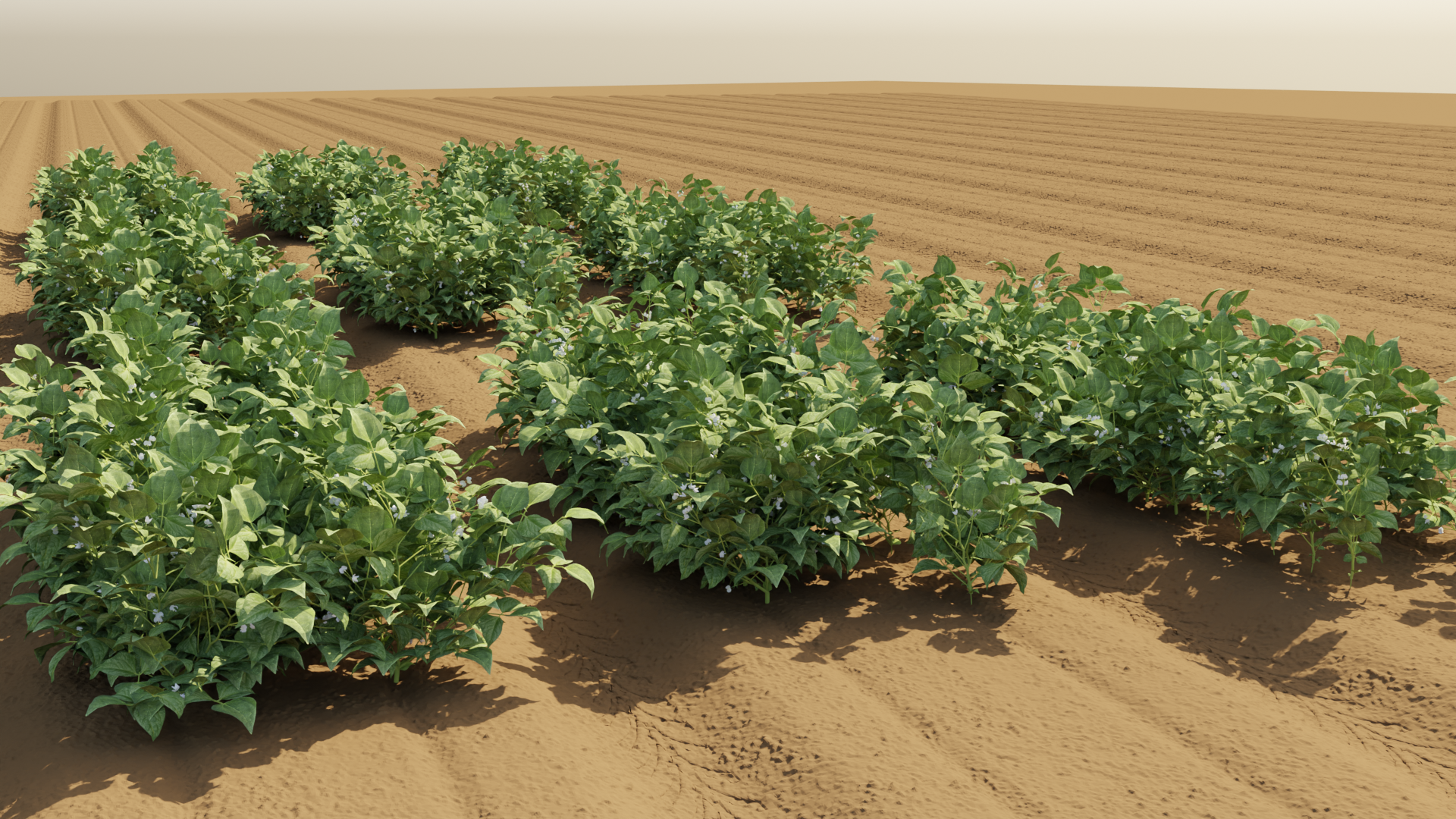}
	\includegraphics[width=0.75\textwidth,trim={0cm 0 6cm 6cm},clip]{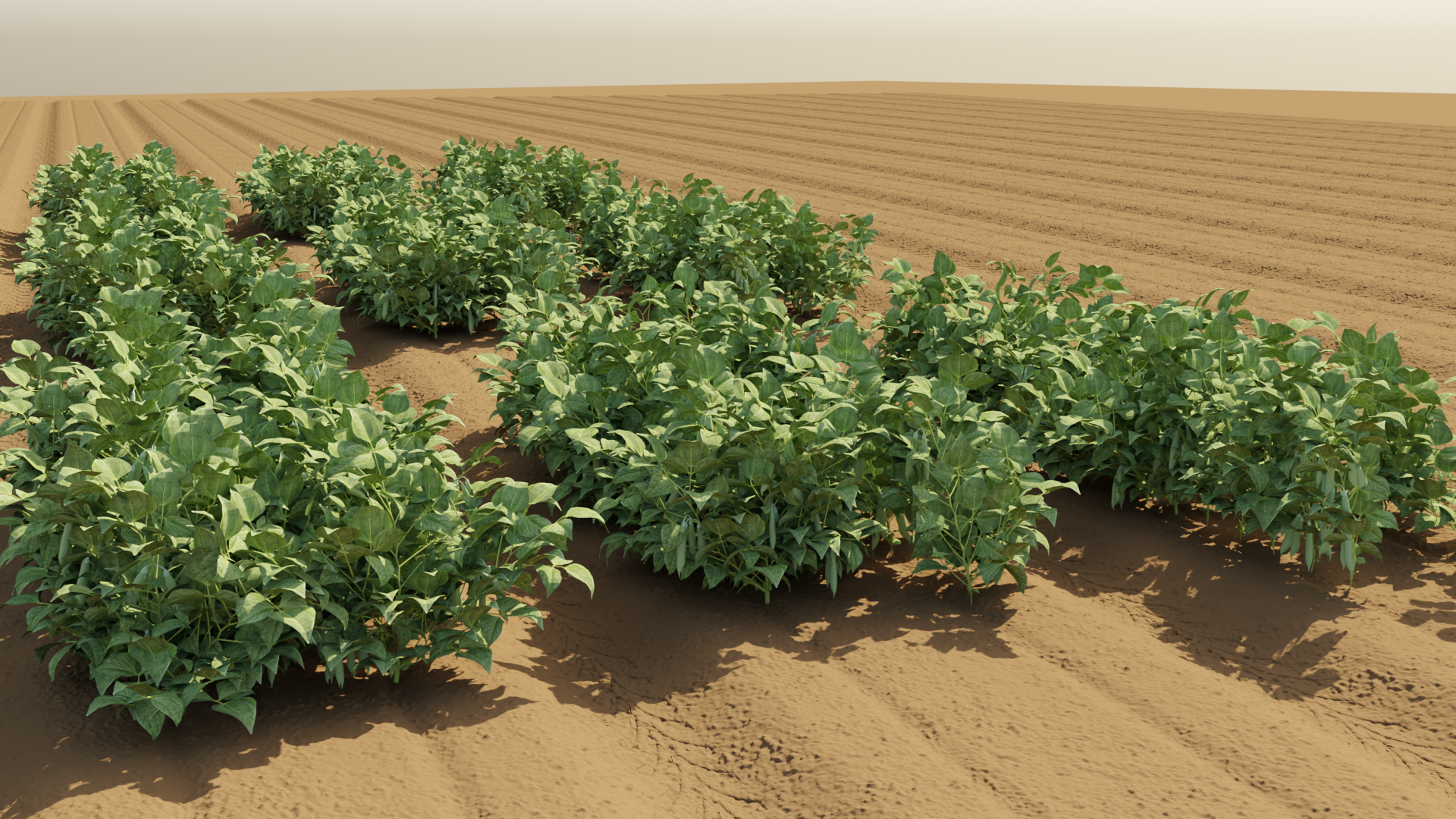}
	\caption{Visualization of common bean model growth over time. (a) 20 days after planting (vegetative growth stage), (b) 50 days after planting (flowering stage), (c) 75 days after planting (pod filling stage).}
	\label{fig:bean_growth}
\end{figure}

\begin{figure}
	\centering
	\includegraphics[width=\textwidth]{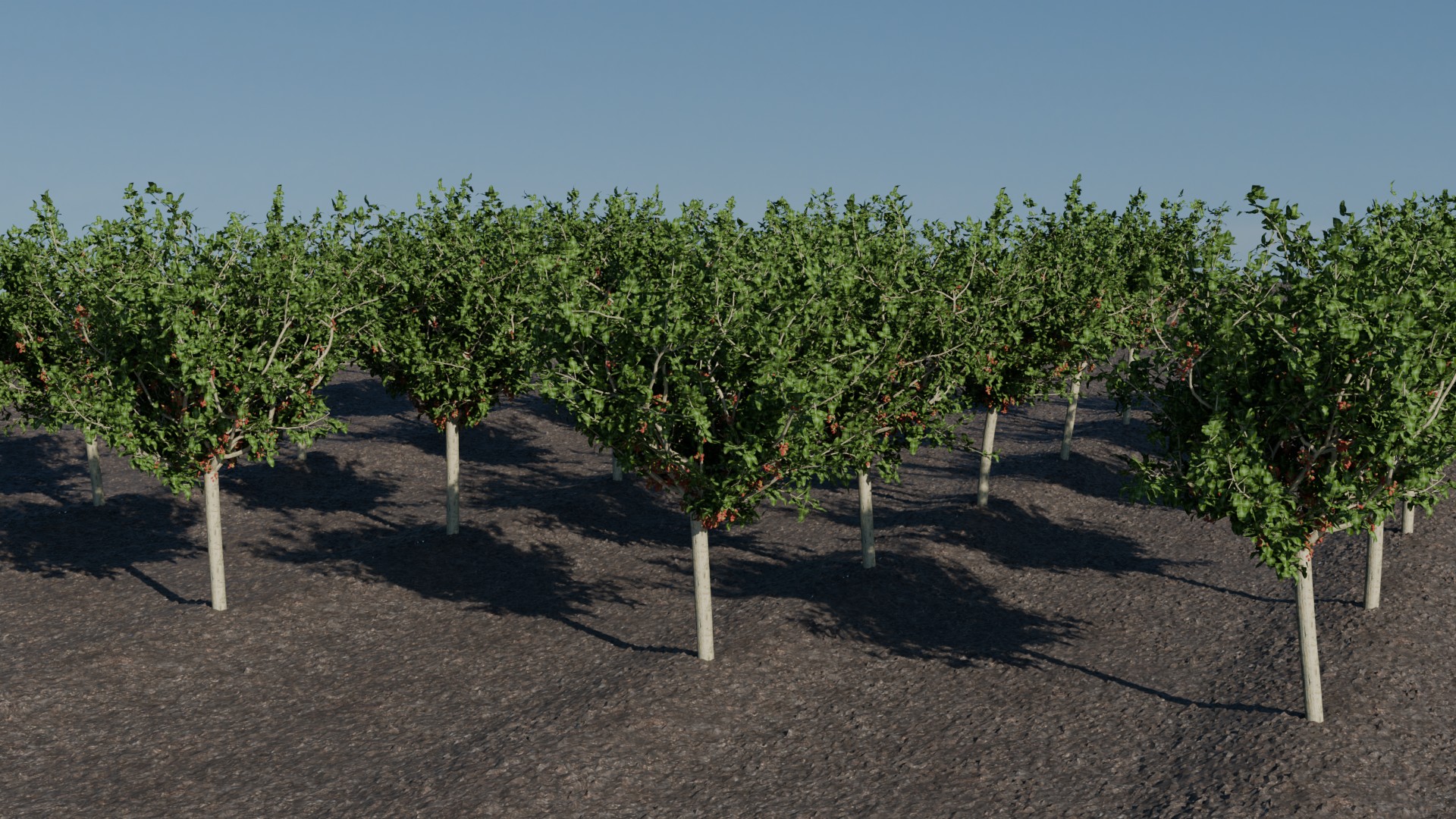}
	\caption{Visualization of a pistachio orchard during reproductive growth stage.}
	\label{fig:pistachio}
\end{figure}

\begin{figure}
	\centering
	\includegraphics[width=\textwidth]{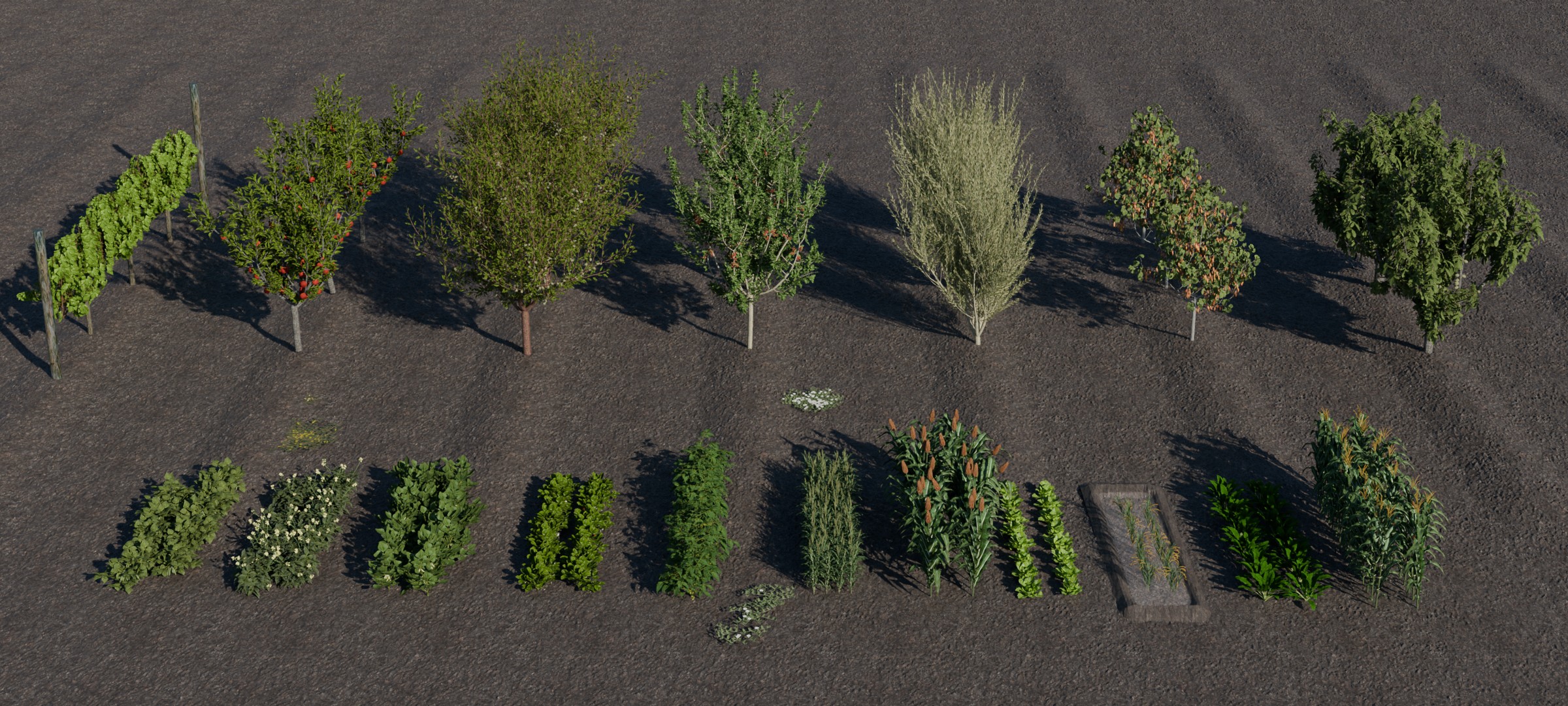}
	\caption{Visualization of 20 different models from the default plant library. Bottom row left-to-right: common bean, cowpea, soybean, strawberry, tomato, wheat, sorghum, lettuce, rice, sugarbeet, maize; Top row left-to-right: grapevine (VSP), apple, almond, pistachio, olive, redbud, walnut. Weed species of puncturevine and bindweed are also shown.}
	\label{fig:manyplants}
\end{figure}

Any plant from the existing library can be loaded based on its associated unique string (cf. second column of Table \ref{tab:plant_library}). Users are recommended to consult the \href{https://plantsimulationlab.github.io/Helios/_plant_architecture_doc.html}{Helios documentation} for the most up-to-date list of available plants in the library. To load a model from the library, users should call \texttt{loadPlantModelFromLibrary()} with a string corresponding to a valid plant model in the library.

Once the model parameters are loaded from the library, users can use \texttt{buildPlantInstanceFromLibrary()} to build an instance of the plant. This builds only a single plant with parameters from the library. Alternatively, users can use \texttt{buildPlantCanopyFromLibrary()} to build a canopy of many plants that are either uniformly spaced or randomly scattered within some defined footprint area. An additional option is available to set an average germination rate, which can allow for introduction of irregularities due to non-uniform germination.

\subsubsection{Modifying parameters of a plant in the existing library}

The default parameters for all plants in the library are defined in the plant architecture plug-in source file \texttt{PlantLibrary.cpp}. These parameters can be modified by manually changing them in the source code, although this is not recommended because it doesn't allow for programmatic control in your code, and they will potentially be overwritten when updating your Helios code version.

To modify parameters of an existing model in the code, users can call \texttt{getCurrentShootParameters()} to get the parameters of the currently loaded plant parameters from the library, and use \texttt{updateCurrentShootParameters()} to update them after making changes. There are two overloaded versions of \texttt{getCurrentShootParameters()}: 1. a string corresponding to one of the shoot types in the model (these are listed in the documentation) is given as input, and the parameters for that shoot type are returned; 2. no input argument is given, and the shoot parameters for all shoot types in the plant are returned in a \texttt{std::map} structure. Example usage is given in Listing \ref{lst:manual_plant_creation}. In either case, the user needs to know the label of the shoot type to be modified, as a given plant may have multiple shoot types. These are listed in the documentation, or users can call \texttt{listShootTypeLabels()} to return a list of label strings contained in a given plant.

\begin{codelisting}
\caption{Sample Helios C++ program to modify parameters of a plant loaded from the library.}
\label{lst:manual_plant_creation}
%\vspace{-12pt}
\begin{lstlisting}[style=cppstyle]
#include "PlantArchitecture.h"
using namespace helios;

int main() {
	// define the Helios Context class (main model database)
	Context context;

	// define an instance of the plant architecture model
	PlantArchitecture plantarchitecture(&context);

	// load "bean" model parameters from the library
	plantarch.loadPlantModelFromLibrary("bean");

	// modify the parameters
	// the "bean" model has shoot types of "unifoliate" and "trifoliate"
	ShootParameters params_uni = plantarch.getCurrentShootParameters("unifoliate");
	params_uni.max_nodes = 20;
	params_uni.phytomer_parameters.internode.pitch = 5.0;
	plantarch.updateCurrentShootParameters("unifoliate", params_uni);
	ShootParameters params_tri = plantarch.getCurrentShootParameters("trifoliate");
	params_tri.max_nodes = 25;
	plantarch.updateCurrentShootParameters("trifoliate", params_tri);

	// add the plant with its base at the origin
	// plant will be built based on updated parameters
	vec3 plant_origin(0,0,0);
	float age = 15; //days
	uint plantID = plantarch.buildPlantInstanceFromLibrary(plant_origin, age);

	return EXIT_SUCCESS;
}
\end{lstlisting}
\end{codelisting}

\subsubsection{Adding a new plant to the library}

To add a completely new plant to the library, the following general workflow is recommended:

\begin{enumerate}
	\item Create any new assets that may be needed. These include creating leaf texture images or 3D models of flower/fruit organs in 3rd party software. Alternatively, it may be possible to use existing assets from other plants in the library if they have acceptable similarity.
	\item Write functions for initializing model parameters and building the initial base shoot (which may just be a single, un-elongated phytomer). It is recommended to start from an existing plant in the library that is most similar to the new plant, and copy-pasting their initialization code. 
    \item Register the initialization functions inside the \texttt{initializePlantModelRegistrations()} method.
    \item Adjust references to assets (e.g., file paths) to any assets created above. Tweak parameters as needed to achieve the desired plant structure.
\end{enumerate}

The documentation gives a more detailed description of each of the steps above.

\subsection{Procedurally growing plant structure over time}\label{S:growing}

Once the initial plant state has been created, either manually or by loading a plant from the library, the model can be grown over time using the \texttt{advanceTime()} method. The user can specify the amount of time in days to advance the model. It is generally more computationally efficient to advance across larger time steps if possible, rather than day-by-day in a loop for example. If there are multiple plants in the scene, all plants are advanced together.

\subsection{Manually building a plant structure}

Methods are available that allow for manually building a plant structure branch-by-branch, which could be used to create a static plant geometry or as an initial structure that could be grown further over time (see Sect. \ref{S:growing}). Manually building the plant structure may be useful in a number of applications such as if the plant is highly constrained (e.g., by a trellis/training system), or if the skeleton of an actual plant is measured and the user wants to use it as the basis for creating a full 3D model. 

Manual creation of a plant always begins with a call to the method \texttt{addPlantInstance()}, which sets the location of the plant and returns a unique ID that can be used to later reference the plant (see Listings \ref{lst:manual_plant_creation}, \ref{lst:manual_plant_creation_python}; C++ and python, respectively). The plant shoot structure is initiated by calling \texttt{addBaseStemShoot()}, which has arguments that define the number of phytomers, base rotation, internode scale, and shoot parameters. At this point new phytomers can be manually appended to the end of the base shoot, other shoot types can be appended, and axillary child shoots can be added at any existing phytomer node. Alternatively, the manually created base structure can be used as an initial state that can be dynamically advanced over time based on the shoot growth parameters that were defined. Users can consult the documentation for more comprehensive details on available methods for manually building the structure.

\begin{codelisting}
	\caption{Sample Helios C++ program for manually building a plant structure}
	\label{lst:manual_plant_creation}
	%\vspace{-12pt}
	\begin{lstlisting}[style=cppstyle]
#include "PlantArchitecture.h"
using namespace helios;

int main() {
	// define the Helios Context class (main model database)
	Context context;

	// define an instance of the plant architecture model
	PlantArchitecture plantarchitecture(&context);

	// Load an existing set of shoot parameters from the library
	// Alternatively, we could manually define a fresh set ourselves
	plantarchitecture.loadPlantModelFromLibrary("bean");

	// add the plant with its base at the origin
	vec3 base_position(0,0,0);
	float age = 15; //days
	uint plantID = plantarchitecture.addPlantInstance(base_position, age);

	return EXIT_SUCCESS;
}
	\end{lstlisting}
\end{codelisting}

\begin{codelisting}
	\caption{Sample PyHelios python program for manually building a plant structure}
	\label{lst:manual_plant_creation_python}
	%\vspace{-12pt}
	\begin{lstlisting}[style=pythonstyle]
from pyhelios import Context, PlantArchitecture
from pyhelios.wrappers.DataTypes import *

# Define the Helios Context (main model database)
context = Context()

# Define an instance of the plant architecture model
plantarchitecture= PlantArchitecture(context)

# Load an existing set of shoot parameters from the library
# Alternatively, we could manually define a fresh set ourselves
plantarchitecture.loadPlantModelFromLibrary("bean")

# Add the plant with its base at the origin
plantID = plantarchitecture.addPlantInstance(base_position=vec3(0,0,0), current_age=15)
	\end{lstlisting}
\end{codelisting}

\subsection{XML file input/output}

L-Systems plant models pioneered the clever approach of encoding plant structure into a string data structure, which enabled efficient representation of model topology in the software and for input/output. There are some drawbacks to this approach, for example that as the model becomes complex with many parameters, the string becomes quite complex and difficult for humans to read. It is also focused on topology, and does not contain the full suite of information to uniquely define all aspects of the plant. 

In the present model, plant topology is represented using a modern object-oriented structure that eliminates the need for a string-based data representation. For file input/output, the model uses the XML file format, which is a hierarchical format similar to HTML that uses human-readable tags to delineate data groups. The hierarchical format lends itself well to the representation of plant topology, and the tag-based delineation improves human readability.

The hierarchical levels of plant representation are plant, shoot, phytomer, petiole, and leaf.  If flowers or fruit are present, the phytomer can also contain peduncles, which can contain flowers/fruit. Each level may contain its own parameters, and may contain one or more instances of the element level below it (except that leaves, flowers, and fruit are the terminal element type). For example, a shoot may contain multiple phytomers, phytomers may contain multiple petioles, and petioles may contain multiple leaves.

\begin{codelisting}
\caption{Sample XML structure for plant topology representation.}
%\vspace{-12pt}
\begin{lstlisting}[style=htmlstyle]
<helios>
  <plant_instance ID="0">
    <base_position> 0 0 0 </base_position>
    <plant_age> 2 </plant_age>
    <shoot ID="0">
      <shoot_type_label> unifoliate </shoot_type_label>
      <parent_shoot_ID> -1 </parent_shoot_ID>
      <parent_node_index> 0 </parent_node_index>
      <parent_petiole_index> 0 </parent_petiole_index>
      <base_rotation> 7.79857 143.494 62.5022 </base_rotation>
      <phytomer>
        <internode>
          <internode_length>0.0063</internode_length>
          <internode_radius>0.000657844</internode_radius>
          <internode_pitch>0</internode_pitch>
          <phyllotactic_angle>213.865</phyllotactic_angle>
          <petiole>
            <petiole_length>0.00623</petiole_length>
            <petiole_radius>0.00145</petiole_radius>
            <petiole_pitch>58.0717</petiole_pitch>
            <petiole_curvature>-94.6669</petiole_curvature>
            <leaflet_scale>1</leaflet_scale>
            <leaf>
              <leaf_scale>0.0084</leaf_scale>
              <leaf_pitch>5.60887</leaf_pitch>
              <leaf_yaw>0</leaf_yaw>
              <leaf_roll>-15</leaf_roll>
            </leaf>
          </petiole>
        </internode>
      </phytomer>
    </shoot>
  </plant_instance>
</helios>
\end{lstlisting}
\label{lst:xml_structure}
\end{codelisting}

\subsection{Using the Helios "project builder" plug-in GUI}

The Helios "project builder" plug-in provides a graphical interface for generating plant models with the plant architecture model, and exporting them to file. Users can select plants available in the library, and set parameters defining the canopy (e.g., plant spacing, age).  If desired, plants can then be exported to PLY (Stanford polygon) or OBJ (Wavefront) file formats from the GUI. Plants in the canopy can be exported together, or automatically one-by-one to separate files.

\section{Future road map}

\subsection{Carbohydrate model}

The plant architecture model currently has a framework implemented for a generalized carbohydrate budget model that couples with plant development. The model tracks the carbon budget on a per-shoot basis, where each shoot has its own carbon pool that can receive carbon through photosynthesis of its leaves or from downstream shoots, and can lose carbon to growth sinks and maintenance respiration. Carbon can limit growth by increasing the phyllochron, by causing flower or fruit abortion, or by causing entire shoot death. 

Initial development has focused on integration of the carbohydrate model with the almond tree model in the library. However, the model is generic and works with any plant model, provided that input parameters are specified appropriately (e.g., carbon costs of leaf, flower, fruit growth respiration). A future publication will describe this model and its validation in detail, and future work will involve calibrating and testing the model for a broader range of species.

\subsection{Nitrogen model}

Work has begun to implement a nitrogen model within the plant architecture model, which couples to other model plug-ins. Nitrogen is tracked across various pools within the plant, including a dynamic pool for each leaf and fruit (if present). Leaves accumulate nitrogen over time according to their local light integral. The leaf-level nitrogen concentration can then be used to calculate pigment concentrations in the leaf optics model plug-in, which are then used to calculate reflectance and transmittance spectra via the PROSPECT model. The leaf-level nitrogen concentration can also be used to calculate physiological model parameters like $V_{cmax}$ in the photosynthesis model. 

An initial version of this functionality is under development, along with its coupling with other plug-ins, and will be fully released in future Helios versions.

\subsection{Automatic calibration from field data}

We are currently developing machine-learning-based approaches to automatically determine parameters in the plant architecture model from field imagery data. Language models can be trained to associate simulated plant renderings with the XML encoding of the corresponding plant structure (Listing \ref{lst:xml_structure}). This can potentially be applied to real imagery data to estimate plant architecture model parameters for plants in the field.

\subsection{Shoot deformation physics}

Future work will add sub models for shoot deformation based on flexible rod bending equations. Representing branch deformation under load is important for representing plant structure under heavy fruit load, or in simulating robotic harvesting and the associated mechanical manipulation of shoots.

\section*{Acknowledgments}

First and foremost, the author would like to acknowledge members of the lab that contributed ideas that went into the development of the modeling framework: Eric Kent, Kyle Rizzo, Ethan Frehner, Ioannis Droutsas, and Heesup Yun. This work was indirectly funded by the Almond Board of California grant HORT.45, the U.S. National Science Foundation grant IOS 2047628, United States Department of Agriculture National Institute of Food and Agriculture, Hatch project number 7003146, and the Bill and Melinda Gates Foundation Project ID: INV-002830. Helpful comments on the manuscript by Dr. Theodore DeJong are also gratefully acknowledged.

%\bibliographystyle{elsarticle-harv}
%\bibliography{Master_Ref}

\newpage

\section*{Appendix: Example visualizations}\label{appendix}

\renewcommand{\thefigure}{A\arabic{figure}}
\setcounter{figure}{0}

\begin{figure}[htb]
	\centering
	\includegraphics[width=0.8\textwidth]{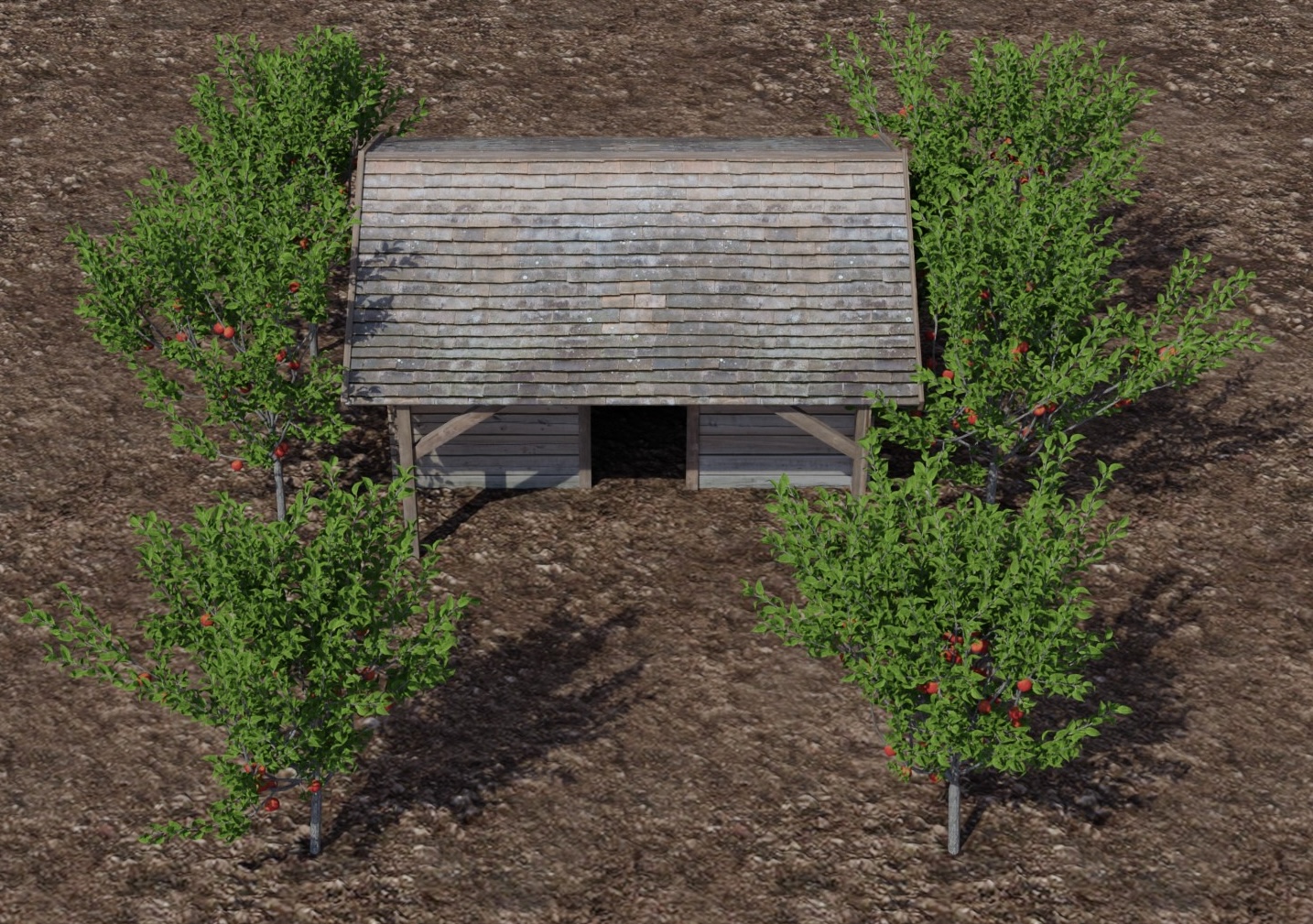}
	\caption{Example visualization illustrating "soft" and "hard" collision avoidance. Trees utilize "soft" collision avoidance to fill open space in the canopy during growth, and utilize "hard" collision avoidance to avoid growing into the house.}
	\label{fig:example_visualization_apple_house}
\end{figure}

\begin{figure}[htb]
	\centering
	\includegraphics[width=\textwidth]{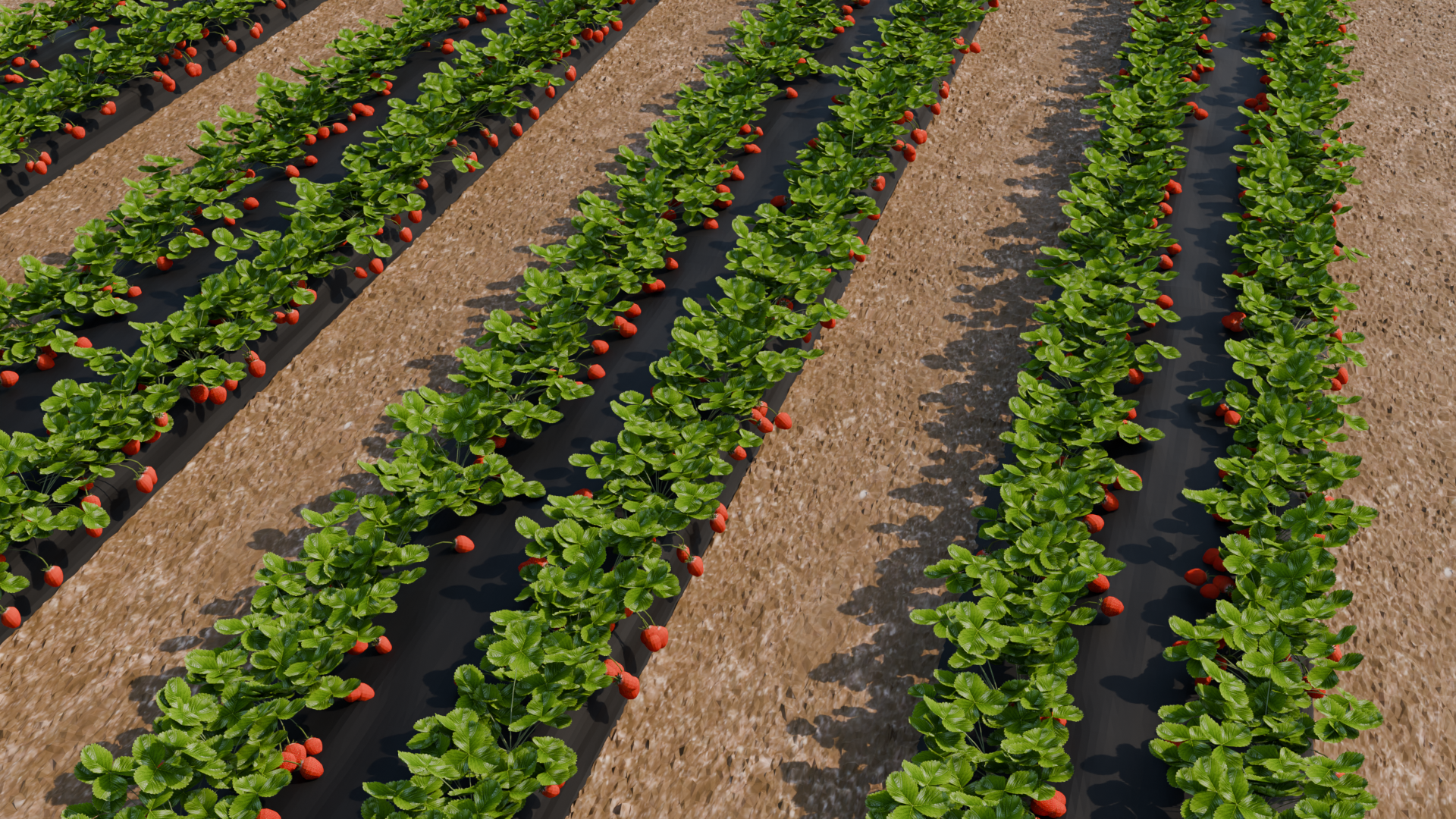}
	\caption{Example visualization illustrating boundary collision avoidance of strawberry fruit. The strawberry plants grow based on constraints from "soft" and "hard" collision avoidance, and fruit are adjusted to avoid ground collisions.}
	\label{fig:example_visualization_strawberry}
\end{figure}

\begin{figure}[htb]
	\centering
	\includegraphics[width=\textwidth]{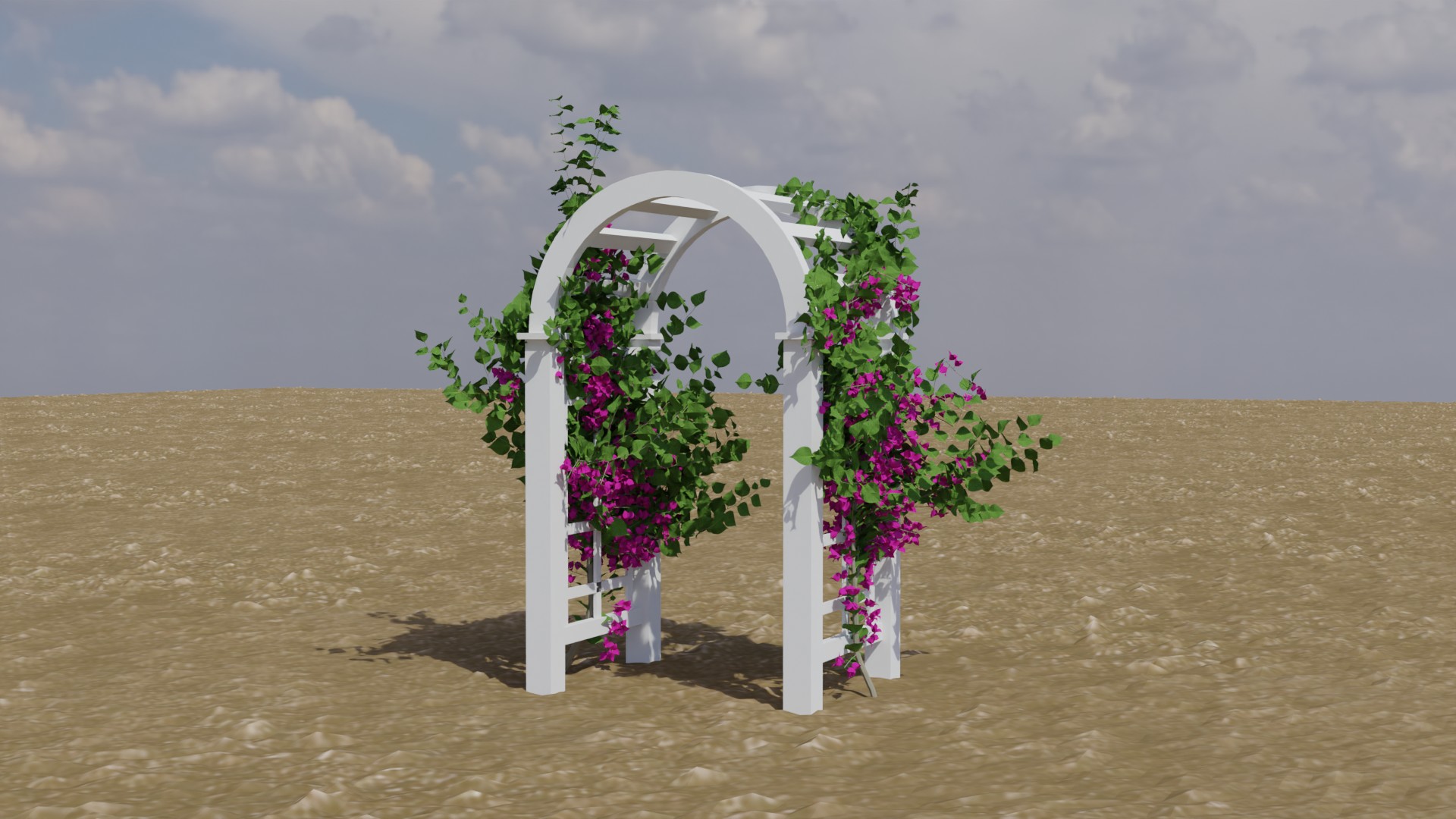}
	\caption{Example visualization illustrating constrained growth of a bougainvillea plant along a trellis based on attraction points.}
	\label{fig:example_visualization_bougainvillea}
\end{figure}

\end{document}